\documentclass[12pt]{article}
\usepackage{graphicx}
\usepackage{booktabs}
 \usepackage{lscape}
 \usepackage{verbatim}
 \usepackage{multirow}
 \usepackage{geometry}
 \usepackage{amssymb}
 \usepackage{units}
 \usepackage{setspace}
 \usepackage{dsfont}
 \usepackage[normalem]{ulem}
 \useunder{\uline}{\ul}{}
\newcommand{\uu}{\underline}
   \usepackage{bm}
   \usepackage{color}
   \usepackage{amsmath}
   \usepackage{comment}
   \usepackage{graphicx,wrapfig,lipsum,multirow,subcaption}
  \usepackage{hyperref}  
      \usepackage[nolists,tablesonly]{endfloat}

\title{Joint TITE-CRM for Dual Agent Dose Finding Studies}

\author{Helen Barnett$^{1}$, Oliver Boix$^2$, Dimitris Kontos$^2$, Thomas Jaki$^{3,4}$\\
$^1$ Lancaster University\\
$^2$ Bayer AG\\
$^3$ MRC Biostatistics Unit, University of Cambridge\\
$^4$ University of Regensburg}
\begin{document}
\maketitle
\begin{abstract}
Dual agent dose-finding trials study the effect of a combination of more than one agent, where the objective is to find the Maximum Tolerated Dose Combination (MTC), the combination of doses of the two agents that is associated with a pre-specified risk of being unsafe. In a Phase I/II setting, the objective is to find a dose combination that is both safe and active, the Optimal Biological Dose (OBD), that optimizes a criterion based on both safety and activity. Since Oncology treatments are typically given over multiple cycles, both the safety and activity outcome can be considered as late-onset, potentially occurring in the later cycles of treatment. This work proposes two model-based designs for dual-agent dose finding studies with late-onset activity and late-onset toxicity outcomes, the Joint TITE-POCRM and the Joint TITE-BLRM. Their performance is compared alongside a model-assisted comparator in a comprehensive simulation study motivated by a real trial example, with an extension to consider alternative sized dosing grids. It is found that both model-based methods outperform the model-assisted design. Whilst on average the two model-based designs are comparable, this comparability is not consistent across scenarios.

\end{abstract}

\section{Introduction} \label{sec:intro}
A traditional Phase I dose-finding study uses binary observations of dose-limiting toxicities (DLTs), evaluating the safety of a novel agent by escalating dose assignment in order to find the Maximum Tolerated Dose (MTD), a dose level associated with a pre-specified probability of observing a DLT. It may however be of interest to study the effects of a combination of more than one agent, for example for synergistic effects of activity or safety. Here the objective is to find the Maximum Tolerated Dose Combination (MTC), the combination of doses of each of the agents associated with the pre-specified probability of observing a DLT.  Multiple designs exist for the dual agent Phase I dose-finding trial, for example the Bayesian Optimal Interval design \cite{lin2017} (BOIN), Keyboard \cite{pan2020}, Partial Order Continual Reassessment Method \cite{wages2011} (POCRM), Waterfall \cite{zhang2016} and Bayesian Logistic Regression Model \cite{Neuenschwander2008} (BLRM), with a comprehensive comparison of model-assisted designs compared to model-based designs recently undertaken \cite{Barnett2024a}.

It is often the case in oncology trials that there are multiple cycles of treatment, and therefore the trial duration would be infeasibly long if one had to wait until the entire follow-up period for the previous cohort had been completed before assigning the next cohort to a dose. Therefore, it is useful for a design to be able to incorporate delayed onset outcomes when only partial information is available, for example on the first cycle out of a three-cycle treatment program. Few designs exist that incorporate delayed onset outcomes in dual agent trials, one such example being the time-to-event version of the POCRM \cite{wages2013}.

In the cases where it is possible to collect measures of activity of the agents, as is often the case in oncology, seamless Phase I/II designs are possible since activity acts as a surrogate for efficacy. Here, the objective is to find the Optimum Biological Dose (OBD), a dose that is both safe and shows sufficient activity that optimizes some utility criterion based on safety and activity, in line with the goal of the FDA's Project Optimus \cite{Optimus} `to move forward with a dose-finding and dose optimization paradigm across oncology that emphasizes selection of a dose or doses that maximizes not only the efficacy of a drug but the safety and tolerability as well.' A small number of designs exist that can evaluate both safety and activity with delayed-onset binary outcomes in the single agent setting, model-assisted approaches such as that from Liu et al \cite{Liu2016} the TITE-B \cite{Yan2019}, and model-based approaches of the $A_T$/$A_E$ Design \cite{Yuan2009} and the Joint TITE-CRM \cite{Barnett2024}. The inclusion of both safety and activity outcomes in dual agent trials with no consideration for late-onset outcomes has been used in various designs (see e.g. \cite{wages2014} \cite{shimamura2018} \cite{yada2018bayesian},\cite{Zhang2023} and \cite{Jaki2024}). However, there does not currently exist a design that considers both late onset toxicity and late onset activity outcomes in the dual agent setting. 

In this work, we propose two model-based approaches to design a seamless Phase I/II dose finding trial in oncology, incorporating delayed-onset toxicity and delayed-onset activity outcomes for dual agents, and compare to a time-to-event extension of the model-assisted BOIN design. A motivating example is given in Section~\ref{sec:mot}, before Section \ref{sec:method} introduces the methodology for all three designs considered. Section \ref{sec:sims} conducts a simulation study to compare the performance of the designs before concluding with a discussion in Section~\ref{sec:dis}.

\section{Motivating Example} \label{sec:mot}
Combining several anticancer agents can enhance overall antitumor activity; however, it may also lead to increased toxicity. Combination therapies are frequently evaluated during the early clinical development of investigational oncology agents. This is particularly relevant for prostate cancer, which remains the most commonly diagnosed cancer in males and a leading cause of cancer-related deaths.

Target Radionuclide Therapy (TRT), which involves therapeutic agents based on radionuclides, holds significant promise in oncology. These new agents are also being investigated in combination with standard treatments for prostate cancer, such as hormone therapy. This approach is especially appealing and more easily planned when both agents are part of the internal portfolio, as preclinical data can be generated more readily, creating a commercially viable opportunity.

This exploration is driven by previous initiatives within Bayer's Targeted Alpha Therapy (TAT) platform. TAT is an emerging modality in the field of radiotherapy that combines tumor-targeting molecules with alpha particle-emitting radioisotopes, aiming to provide a novel approach to cancer treatment and potentially overcome resistance \cite{Bidkar2024}.

In an early-phase study, a thorium-227 labelled antibody-chelator conjugate was combined with darolutamide, a synthetic nonsteroidal next-generation androgen receptor antagonist, in patients with metastatic castration-resistant prostate cancer \cite{Example_trial227, Example_trial4}. In this context, the doses of both agents could be adjusted to identify the optimal dose combination that enhances overall antitumor activity while maintaining an acceptable safety profile.

\section{Methodology} \label{sec:method}
We propose two alternative approaches, both based on the Joint TITE-CRM \cite{Barnett2024}, a design for single agent therapies with late-onset activity and toxicity outcomes, which was shown to have good operating characteristics. The first proposed approach, the Joint TITE-POCRM, extends the Joint TITE-CRM to the dual agent setting by mapping the two-dimensional dosing grid into one dimension using the Partial Order Continual Reassessment Method (POCRM) \cite{wages2011}. The second, the Joint TITE-BLRM, extends the Joint TITE-CRM to the dual agent setting by modelling the joint odds using the Bayesian Logistic Regression Model (BLRM) \cite{neuenschwander2015}. These are compared to the model assisted TITE-BOIN12 design \cite{Zhou2022}, which we extend to the dual agent setting, summarised in this section, with further details given in the Supplementary Materials. Section~\ref{sec:setup} provides the trial setup that is applicable to all methods considered, then the following subsections introduce each of the approaches in turn. For brevity in this section, many equations that are identical in the cases of toxicity and activity are only given once, with relevant subscripts of $T$ for toxicity that can be replaced with $A$ in the case of activity.

\subsection{Trial Setup} \label{sec:setup}
\subsubsection{Dosing Grid}
We consider two agents labelled agent $W_1$ and agent $W_2$; agent $W_1$ has $I$ dose levels labelled $i=1, \ldots , I$ and agent $W_2$ has $J$ dose levels labelled $j=1, \ldots , J$. The dose combinations are therefore labelled $d_{ij}$, making up an $I$ by $J$ dosing grid, with an example of such a grid for $I=2$ and $J=5$ shown in Table \ref{tab:example}.

\begin{table}[h!]
\begin{tabular}{lllllll}
                                              &                           & \multicolumn{5}{c}{Agent $W_2$}                                                                                                                                   \\ \cline{3-7} 
                                              & \multicolumn{1}{l|}{}     & \multicolumn{1}{l|}{50kBq/kg}     & \multicolumn{1}{l|}{75kBq/kg}     & \multicolumn{1}{l|}{100kBq/kg}     & \multicolumn{1}{l|}{125kBq/kg}     & \multicolumn{1}{l|}{150kBq/kg}     \\ \cline{2-7} 
\multicolumn{1}{c|}{\multirow{2}{*}{Agent $W_1$}} & \multicolumn{1}{l|}{600mg} & \multicolumn{1}{l|}{$d_{11}$} & \multicolumn{1}{l|}{$d_{12}$} & \multicolumn{1}{l|}{$d_{13}$} & \multicolumn{1}{l|}{$d_{14}$} & \multicolumn{1}{l|}{$d_{15}$} \\ \cline{2-7} 
\multicolumn{1}{c|}{}                         & \multicolumn{1}{l|}{1200mg} & \multicolumn{1}{l|}{$d_{21}$} & \multicolumn{1}{l|}{$d_{22}$} & \multicolumn{1}{l|}{$d_{23}$} & \multicolumn{1}{l|}{$d_{24}$} & \multicolumn{1}{l|}{$d_{25}$} \\ \cline{2-7} 
\end{tabular}
\caption{An example of dose labelling for the dual agent trial.\label{tab:example}}
\end{table}
\subsubsection{Trial Procedure} \label{sec:procedure}
\begin{figure}[h!]
  \centering
  \includegraphics[width=1\linewidth]{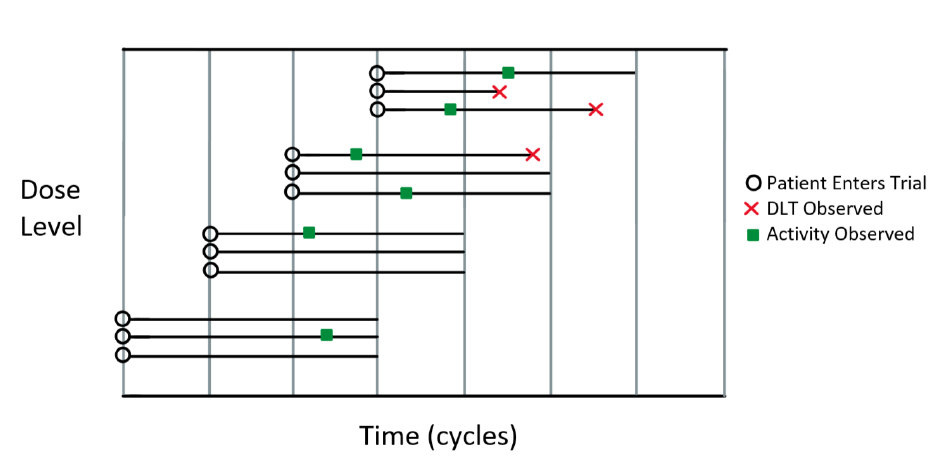}
\caption{Visualisation of example trial with an observation period of 3 cycles and dose-escalation decisions taking place after 1 cycle.}
\label{fig:Ex}
\end{figure}
Patients are followed up for a total of $\tau$ treatment cycles, with the trial proceeding as follows:
\begin{enumerate}
\item The first cohort of patients is treated at the pre-specified starting dose combination.
\item After the previous cohort has been observed for one cycle of treatment, the next cohort is assigned to a dose combination based on admissibility and a criterion of utility, both incorporating activity and toxicity.
\item Repeat step 2 until a stopping rule is triggered.
\end{enumerate}
Figure \ref{fig:Ex} illustrates how such a trial would progress, with each vertical line representing the start of a cycle. Patients may observe an activity response only, an activity response and then a DLT response, a DLT response only, or neither an activity nor DLT response, with all four types of outcome visualised here.

\subsubsection{Admissibility and Utility} \label{sec:admiss}
In all designs, a cohort may only be assigned to dose combination $d_{ij}$ if that dose combination is deemed to be admissible. This means that it is above a certain probability of being both safe and active, and is formally defined as:
\[
\mathbb{P} (\pi_T(d_{ij}) < \phi_T) > q_T 
\hspace{15pt} \mbox{\&} \hspace{15pt} 
\mathbb{P} (\pi_A(d_{ij})  > \phi_A) > q_A.
\]
for some pre-defined thresholds $q_T$ for toxicity and $q_A$ for activity, where $\phi_T$ and $\phi_A$ are the target probabilities for toxicity and activity, respectively, for the follow-up period $\tau$, and $\pi_T(d_{ij}) $ and $\pi_A(d_{ij}) $ are the probabilities of toxicity and activity at dose combination $d_{ij}$, respectively. This ensure that assignment is to a dose combination that is considered neither futile nor unsafe.

Once dose combinations are deemed admissible, the dose combination that optimizes some utility criterion is chosen. For both the Joint TITE-POCRM and the Joint TITE-BLRM, the following utility as used by Barnett et al. \cite{Barnett2024} is used:
\begin{equation}
U(\pi_A, \pi_T) = \pi_A - \omega_1 \pi_T- \omega_2 \pi_T \mathds{I}(\pi_T > \phi_T), \label{eq:utility1}
\end{equation}
where the two weights, $\omega_1$ and $\omega_2$,  reflect the trade-off between toxicity and activity. Note that the Joint TITE-POCRM and the Joint TITE-BLRM both use the same utility criterion, but the model used to inform this utility is different.
\subsection{Joint TITE-POCRM}
The Partial Order Continual Reassessment Method (POCRM)\cite{wages2011} maps the two-dimensional dosing grid into one dimension so that a single-agent dose-response model can be applied. In order to do this, one must choose what is known as the dose skeleton, the dose-response model and the ordering of the mapping itself.

Firstly, consider the dose skeleton, $x$. This can be interpreted for many models as the prior probability of observing a toxicity response on a given dose; a vector of increasing values between 0 and 1, with a length equal to the number of doses. This is used in the dose-response model as opposed to the real dose values, with the advantage of being invariant to units of measurement.

Next, consider the dose-response model. In the original proposal of the POCRM, the one parameter power model was used:
\[
F (x , \alpha) = {x}^{\exp(\alpha)},
\]
where $\alpha$ is the parameter of the model, assigned a $N(0,1.34^2)$ prior. Although displaying reasonable operating characteristics, the two parameter logistic model allows for much more flexibility in the dose response relationship:
\[
F(x,\bm{\beta})= \frac{\exp(\beta_{0} + \beta_{1} x)}{1+\exp(\beta_{0} + \beta_{1} x)},
\]
where $\bm{\beta}=(\beta_0, \beta_1)$ is the parameter vector, with $ (\beta_0, \log(\beta_1))$ being assigned a bivariate normal prior, $\begin{pmatrix}
\beta_0 \\
\log (\beta_1)
\end{pmatrix} \sim N_2 \left( \begin{pmatrix}
c_1\\
c_2
\end{pmatrix} , 
\begin{pmatrix}
v_1 & 0 \\
0 & v_2 
\end{pmatrix} \right)$. 

In this setting, since responses may be late-onset, we also consider the time-to-event of a response in the model:
\[
G(x, \cdot, w) = w F(x, \cdot),
\]
where $w$ is a function of the time-to-event of a response.

Next, consider the mapping. In order to choose which dose combinations are mapped to which elements of the skeleton, one has to consider the ordering of the dose combinations. For example, one may map the grid in Table \ref{tab:example} into one dimension using the ordering:
\[
R(d_{11}) \leq 
R(d_{21}) \leq 
R(d_{12}) \leq 
R(d_{22}) \leq 
R(d_{13}) \leq 
R(d_{23}) \leq 
R(d_{14}) \leq 
R(d_{24}) \leq 
R(d_{15}) \leq 
R(d_{25}) 
\]
where $R(d_{ij})$ represents the toxicity response rate of dose $d_{ij}$. This is just one of many possible orderings. Using the dose skeleton of $(0.15, 0.19, 0.22, 0.26, 0.30, 0.34, 0.38, 0.42, 0.46,  0.50)$ would map $d_{11}$ to 0.15, $d_{21}$ to 0.19 and so on.

A one-dimensional dose-response model can then be applied. Since there are a large number of potential mappings that can exist, the POCRM considers only a subset of these, based on the ordering of the dosing grid. To choose the one dimensional ordering in which to apply the dose skeleton to the dosing grid, a set of so-called partial orderings are considered. These are plausible orderings of the dose-combinations that adhere to the dose orderings. For example, increasing the dose level of one agent whilst keeping the other the same would be a "higher" dose. (e.g. $R(d_{12}) \geq R(d_{11})$) However, when increasing the dose level of one agent and decreasing the other (an off-diagonal move) one would not necessarily be able to specify which dose combination is "higher" (e.g. $R(d_{12}) \geq R(d_{21)}$ or $R(d_{12}) \leq R(d_{21})$). Further restrictions may be placed on the set of partial orderings in order to define the subset to consider. These restrictions may be drawn from information about the agents themselves, or from a statistical perspective. It is worth noting that increasing the number of considered partial orderings substantially increases the computational cost of the design, and so including  multiple orderings that are very similar may not yield any increase in performance. In general, one must consider enough orderings to capture the main differences that could occur, as originally argued by Wages et al \cite{wages2011} in the suggestion of "good" orderings.

In this application, since we are considering both activity and toxicity endpoints, how they are incorporated into each part of the POCRM must be taken into account. The skeleton is not necessarily the same for activity and toxicity, but is fixed for each prior to conducting the study. The dose-response model must be for both toxicity and activity and the mapping must be done for activity and toxicity separately. Like the dose skeleton, the set of partial orderings is not necessarily the same for activity and toxicity, and hence the chosen ordering is also not necessarily the same.

The Joint TITE-POCRM therefore breaks down into two parts at each dosing decision:
\begin{itemize}
\item \textbf{Part 1: } Choose the partial ordering for toxicity and the partial ordering for activity with the highest associated posterior probabilities to be used for model fitting.
\item \textbf{Part 2:} Use the chosen orderings for toxicity and activity from Part 1 to fit the Joint TITE-CRM. This is used to determine the next dose combination.
\end{itemize}
Since they are two distinct parts, it is not necessary that the models used in each part are the same. The gain in accuracy of using a two-parameter joint logistic model in Part 2 as opposed to a one-parameter independent power model is large, as this gives much more flexibility and allows the most use out of the observed data to estimate this relationship \cite{Barnett2022}. However in Part 1, not much is gained by the more complex model, which consequently substantially increases computational burden. This is because the objective is only to find the right ordering, and not precise estimation. Therefore it is proposed here to use one-parameter independent power models in Part 1 to determine the ordering, and a two parameter joint logistic model in Part 2 to determine the OBD.

\textbf{Part 1: Choosing partial orderings}\\
A set of  $M_T$ partial orderings for toxicity, labelled $m_T=1, \ldots M_T$, and a set of $M_A$ partial orderings for activity, labelled $m_A=1, \ldots M_A$, are considered. The dose-response model for each partial ordering is referred to as a working model. For each set, a prior probability is assigned to each partial ordering within the set, these are $p(m_A)$ and $p(m_T)$ with $\sum_{m_A=1}^{M_A} p(m_A)=\sum_{m_T=1}^{M_T} p(m_T)=1$. This probability represents the prior belief that the corresponding partial ordering is the correct ordering.

Toxicity and activity are considered independently in this part, both following \cite{wages2013}, with the following dose-response models:
\[
G_{P1} (x^{(T)}_{\ell , m_T} , w_{\ell}^{(T)} , \alpha^{(T)}) = w_{\ell}^{(T)} {(x^{(T)}_{\ell , m_T})}^{\exp(\alpha^{(T)})}
\]

\[
L_{P1,m_T}(\alpha^{(T)} |\Omega_{m_T}) = \prod_{\ell=1}^{n} G_{P1} (x^{(T)}_{\ell , m_T} , w_{\ell}^{(T)} , \alpha^{(T)})^{y^{(T)}_{\ell}}  (1-G_{P1} (x^{(T)}_{\ell , m_T} , w_{\ell}^{(T)} , \alpha^{(T)}))^{(1-y^{(T)}_{\ell})}
\]
for up to and including patient $n$, where $x^{(T)}_{\ell ,m_T}$ is the toxicity skeleton value for the dose given to patient $\ell$ under working model $m_T$, $y^{(T)}_{\ell}$ is the binary toxicity outcome for patient $\ell$, $w_{\ell}^{(T)}$ is the time-to-event weight for toxicity for patient $\ell$, $\alpha^{(T)}$ is the parameter of the dose-response model under working model $m_T$. Similarly for activity, replacing $T$ with $A$.

The time-to-event weight for toxicity, $w^{(T)}$, is defined as the proportion of total follow-up, as used by \cite{YingKuenCheung2000}, unless a DLT is observed in which case $w^{(T)}_{\ell}=1$. In a similar fashion, the time-to-event weight for activity, $w^{(A)}$ is defined as the proportion of total follow-up, unless an activity outcome is observed, in which case $w^{(A)}_{\ell}=1$. However, if a DLT is observed before an activity outcome can be observed, then the activity observation is censored by DLT time and so $w^{(A)}_{\ell}=(\mbox{DLT time - entry time})/\mbox{Total follow up time}$. This is the same specification as used by the Joint TITE-CRM \cite{Barnett2024}.

For activity and toxicity, a prior $g$ is elicited on the parameter $\alpha$, and the likelihoods are calculated for each of the considered working models. For each working model, the posterior probability of the corresponding partial ordering for toxicity is therefore calculated as:
\[
p (m_T|\Omega_n)=\frac{p(m_T) \int_{\alpha^{(T)}} L_{m_T} (\alpha^{(T)}| \Omega_n)g(\alpha^{(T)})d\alpha^{(T)}}{\sum_{m_T=1}^{M_T} p(m_T) \int_{\alpha^{(T)}} L_{m_T} (\alpha^{(T)}| \Omega_n)g(\alpha^{(T)})d\alpha^{(T)}} ,
\]
and similarly for activity, again replacing $T$ with $A$. For both activity and toxicity, the respective individual working models with the highest posterior probability, labelled $m^*_A$ and $m^*_T$, are chosen to proceed to Part 2.

\textbf{Part 2: Determining next dose combination}\\
The chosen partial orderings from Part 1 are used in the following model, a joint two-parameter logistic model for both safety and activity outcomes, the Joint TITE-CRM \cite{Barnett2024}. The model for toxicity is outlined here, with the same used for activity, replacing $T$ with $A$:
\[
F(x^{(T)}_{\ell ,m^*_T},\bm{\beta^{(T)}})= \frac{\exp(\beta^{(T)}_{0} + \beta^{(T)}_{1} x^{(T)}_{\ell ,m^*_T})}{1+\exp(\beta^{(T)}_{0}  + \beta^{(T)}_{1} x^{(T)}_{\ell ,m^*_T})},
\]
where $\bm{\beta^{(T)}}= (\beta^{(T)}_{0}, \beta^{(T)}_{1})$. Similarly for activity, replacing $T$ with $A$. \\
The same weights for toxicity and activity observations as used in part 1 ($w_{\ell}^{(T)}$ and $w_{\ell}^{(A)}$ respectively) are used in the weighted dose response models:
\[ 
G_{P2}(x^{(T)}_{\ell ,m^*_T},w_{\ell}^{(T)},\bm{\beta^{(T)}}) = w_{\ell}^{(T)} F(x^{(T)}_{\ell ,m^*_T},\bm{\beta^{(T)}}),
\]

The $G_{P2}$ for activity and toxicity are related by a Gumbel Model with correlation parameter $\psi$ considering all binary combinations of toxicity and activity outcomes, as in the Joint TITE-CRM, with full details given in the Supplementary Materials.

Priors are elicited on the parameters $\psi$, $\beta^{(A)}_0$, $\beta^{(A)}_1$, $\beta^{(T)}_0$, $\beta^{(T)}_1$ using the calibration procedure outlined in the Supplementary Materials, and the likelihood, also in the Supplementary Materials, is used to update the joint posterior for all of the parameters using MCMC methods. The posterior distributions of the probability of activity and toxicity on all cycles at each dose combination, respectively,  $\pi_A$ and $\pi_T$, are used to determine the admissibility and utility of each dose combination, and therefore the recommended next dose combination. 

Out of the admissible dose combinations, the dose combination with the highest utility is recommended. Following the methodology of the Joint TITE-CRM \cite{Barnett2024}, the below utility function is used:
\begin{equation}
U(\pi_A, \pi_T) = \pi_A - \omega_1 \pi_T- \omega_2 \pi_T \mathds{I}(\pi_T > \phi_T). \label{eq:utility}
\end{equation}

\subsection{Joint TITE-BLRM}
The Joint TITE-BLRM uses the actual dose values of agent $W_1$, $d_i^{(W_1)}$ and agent $W_2$, $d_j^{(W_2)}$ in the model. The toxicity of the two agents is modelled, and their odds linked by an odds-multiplier (as recommended by \cite{Neuenschwander2008}), and the same is done for activity. The dual agent time-to-event activity and toxicity models are then linked by a Gumbel model with correlation parameter $\psi$ in a similar fashion to the Joint TITE-CRM.

The model for toxicity is outlined here, with the same used for activity, replacing $T$ with $A$. The toxicity of agent $W_1$ is given as:
\[
F(d_i^{(W_1)},\bm{\beta^{(W_1, T)}})= \frac{\exp(\beta^{(W_1, T)}_{0} + \beta^{(W_1, T)}_{1} d_i^{(W_1)})}{1+\exp(\beta^{(W_1, T)}_{0}  + \beta^{(W_1, T)}_{1} d_i^{(W_1)})},
\]
where $\bm{\beta^{(W_1, T)}}=(\beta^{(W_1, T)}_{0},\beta^{(W_1, T)}_{1})$, and the toxicity of agent $W_2$ is given as:
\[
F(d_j^{(W_2)},\bm{\beta^{(W_2, T)}})= \frac{\exp(\beta^{(W_2, T)}_{0} + \beta^{(W_2, T)}_{1} d_j^{(W_2)})}{1+\exp(\beta^{(W_2, T)}_{0}  + \beta^{(W_2, T)}_{1} d_j^{(W_2)})},
\]
where $\bm{\beta^{(W_2, T)}}=(\beta^{(W_2, T)}_{0},\beta^{(W_2, T)}_{1})$, with their odds linked:
\[
odds^{(T)}_{d_i^{(W_1)},d_j^{(W_2)}}=odds^{(T),0}_{d_i^{(W_1)},d_j^{(W_2)}} \exp( \eta^{(T)} d_i^{(W_1)}d_j^{(W_2)}),
\]
where
\begin{align*}
odds^{(T),0}_{d_i^{(W_1)},d_j^{(W_2)}}=\frac{F(d_i^{(W_1)},\bm{\beta^{(W_1, T)}})}{1-F(d_i^{(W_1)},\bm{\beta^{(W_1, T)}})} &+ \frac{F(d_j^{(W_2)},\bm{\beta^{(W_2, T)}})}{1-F(d_j^{(W_2)},\bm{\beta^{(W_2, T)}})} + \\ &\frac{F(d_i^{(W_1)},\bm{\beta^{(W_1, T)}})F(d_j^{(W_2)},\bm{\beta^{(W_2, T)}})}{(1-F(d_i^{(W_1)},\bm{\beta^{(W_1, T)}}))(1-F(d_j^{(W_2)},\bm{\beta^{(W_2, T)}}))}.
\end{align*}

The contribution to the likelihood is then weighted in the same way as in the Joint TITE-POCRM, with the same definition of weights $w$:
\[ 
G_B(x_{\ell}^{W_1},x_{\ell}^{W_2},w_{\ell}^{(T)},\bm{\beta^{(W_1,T)}},\bm{\beta^{(W_2,T)}}, \eta^{(T)}) = w_{\ell}^{(T)} \frac{odds^{(T)}_{x_{\ell}^{W_1},x_{\ell}^{W_2}}}{1+odds^{(T)}_{x_{\ell}^{W_1},x_{\ell}^{W_2}}},
\]
where $x_{\ell}^{W_1}$ is the dosage of agent $W_1$ assigned to patient $\ell$, and $x_{\ell}^{W_2}$ is the dosage of agent $W_2$ assigned to patient $\ell$.

There are therefore  11 parameters in this model:
 $\beta_0^{W_2,A}$ , $\beta_1^{W_2,A}$ , $\beta_0^{W_1,A}$ , $\beta_1^{W_1,A}$ , $\beta_0^{W_2,T}$ , $\beta_1^{W_2,T}$ , $\beta_0^{W_1,T}$ , $\beta_1^{W_1,T}$ , $\eta^{(T)}$, $\eta^{(A)}$, $\psi$.  Priors are elicited on all 11 parameters, again using the same calibration procedure as the Joint TITE-POCRM, detailed in the Supplementary Materials and the likelihood also outlined in the Supplementary Materials is used to update the joint posterior for all of the parameters using MCMC methods. In the same way as the Joint TITE-POCRM, the posterior distributions of the probability of activity and toxicity on all cycles at each dose combination, respectively,  $\pi_A$ and $\pi_T$, are used to determine the admissibility and utility of each dose combination, and therefore the recommended next dose combination. The same utility criterion (\ref{eq:utility}) as used for the Joint TITE-POCRM is used for the Joint TITE-BLRM.
 \subsection{TITE-BOIN12}
As a model-assisted design, TITE-BOIN12 \cite{Zhou2022} uses a model at each dose level to make escalation decisions, rather than modelling the dose-response relationship as the previous two designs have. The original proposal of TITE-BOIN12 is for single agent trials, therefore here we have extended the approach in line with the dual agent BOIN design \cite{lin2017}, using the utility criterion of the TITE-BOIN12 and the escalation approach of the dual agent BOIN. Further details of this model-assisted comparator are given in the Supplementary Materials.

\section{Simulations} \label{sec:sims}
In order to assess the operating characteristics of the designs, simulations are conducted over a range of plausible scenarios.

The setting is motivated by the example illustrated in Section \ref{sec:mot}, with two doses of Agent $W_1$ and five doses of Agent $W_2$. The starting dose is $d_{22}$, reflecting the fact that the lowest dose levels of each agent are fall-back doses. The procedure of the trial follows that outlined in Section \ref{sec:procedure}.

Details of the values used in the priors of all methods are given in the supplementary materials.

\subsection{Scenarios}
Given the complexity of the setting, it is important to consider a range of relevant scenarios. Six safety scenarios and six activity scenarios are combined to give 36 total scenarios. The individual safety and activity scenarios are defined so that different areas of the grid are active and safe. When combined, this gives the OBD in different positions in the grid. The six toxicity and activity scenarios are given in Table \ref{tab:scenTA}, with the probability of toxicity and activity for the whole observation window of $\tau=3$ cycles quoted. The individual utility of the 36 scenarios for the two CRM-based methods and the TITE-BOIN12 are provided in the supplementary materials.

\begin{table}[h!]
\begin{tabular}{ll|llllllll|lllll}
\cline{2-7} \cline{10-15}
                                        & \multicolumn{1}{c|}{\multirow{2}{*}{\begin{tabular}[c]{@{}c@{}}Agent $W_1$ \\ (mg)\end{tabular}}} & \multicolumn{5}{c}{Agent $W_2$ (kBq/kg)} &  &                                         & \multicolumn{1}{c|}{\multirow{2}{*}{\begin{tabular}[c]{@{}c@{}}Agent $W_1$ \\ (mg)\end{tabular}}} & \multicolumn{5}{c}{Agent $W_2$ (kBq/kg)} \\ \cline{3-7} \cline{11-15} 
                                        & \multicolumn{1}{c|}{}                                                                         & 50    & 75    & 100   & 125   & 150  &  &                                         & \multicolumn{1}{c|}{}                                                                         & 50    & 75    & 100   & 125   & 150  \\ \cline{2-7} \cline{10-15} 
\multicolumn{1}{c}{\multirow{2}{*}{T1}} & 600                                                                                           & 0.03  & 0.07  & 0.11  & 0.15  & 0.2  &  & \multicolumn{1}{c}{\multirow{2}{*}{A1}} & 600                                                                                           & 0.2   & 0.3   & 0.35  & 0.5   & 0.55 \\
\multicolumn{1}{c}{}                    & 1200                                                                                          & 0.05  & 0.09  & 0.13  & 0.25  & 0.3  &  & \multicolumn{1}{c}{}                    & 1200                                                                                          & 0.25  & 0.4   & 0.45  & 0.6   & 0.65 \\ \cline{2-7} \cline{10-15} 
\multirow{2}{*}{T2}                     & 600                                                                                           & 0.1   & 0.15  & 0.2   & 0.3   & 0.4  &  & \multirow{2}{*}{A2}                     & 600                                                                                           & 0.3   & 0.32  & 0.38  & 0.4   & 0.46 \\
                                        & 1200                                                                                          & 0.45  & 0.5   & 0.55  & 0.6   & 0.6  &  &                                         & 1200                                                                                          & 0.34  & 0.36  & 0.42  & 0.44  & 0.48 \\ \cline{2-7} \cline{10-15} 
\multirow{2}{*}{T3}                     & 600                                                                                           & 0.05  & 0.08  & 0.15  & 0.2   & 0.45 &  & \multirow{2}{*}{A3}                     & 600                                                                                           & 0.06  & 0.08  & 0.12  & 0.2   & 0.3  \\
                                        & 1200                                                                                          & 0.1   & 0.12  & 0.3   & 0.4   & 0.5  &  &                                         & 1200                                                                                          & 0.1   & 0.15  & 0.25  & 0.35  & 0.4  \\ \cline{2-7} \cline{10-15} 
\multirow{2}{*}{T4}                     & 600                                                                                           & 0.1   & 0.2   & 0.4   & 0.5   & 0.6  &  & \multirow{2}{*}{A4}                     & 600                                                                                           & 0.05  & 0.2   & 0.3   & 0.4   & 0.5  \\
                                        & 1200                                                                                          & 0.3   & 0.45  & 0.55  & 0.6   & 0.6  &  &                                         & 1200                                                                                          & 0.1   & 0.25  & 0.35  & 0.45  & 0.55 \\ \cline{2-7} \cline{10-15} 
\multirow{2}{*}{T5}                     & 600                                                                                           & 0.3   & 0.45  & 0.5   & 0.55  & 0.6  &  & \multirow{2}{*}{A5}                     & 600                                                                                           & 0.1   & 0.12  & 0.14  & 0.16  & 0.18 \\
                                        & 1200                                                                                          & 0.4   & 0.5   & 0.55  & 0.6   & 0.6  &  &                                         & 1200                                                                                          & 0.2   & 0.3   & 0.4   & 0.5   & 0.6  \\ \cline{2-7} \cline{10-15} 
\multirow{2}{*}{T6}                     & 600                                                                                           & 0.4   & 0.4   & 0.5   & 0.5   & 0.6  &  & \multirow{2}{*}{A6}                     & 600                                                                                           & 0.1   & 0.1   & 0.1   & 0.1   & 0.1  \\
                                        & 1200                                                                                          & 0.4   & 0.4   & 0.5   & 0.5   & 0.6  &  &                                         & 1200                                                                                          & 0.1   & 0.1   & 0.1   & 0.1   & 0.2  \\ \cline{2-7} \cline{10-15} 
\end{tabular}
\caption{Scenarios considered for simulation study, giving the probability of observing a toxicity/activity event at each dose combination. Six for safety, labelled T1-T6, six for activity labelled A1-A6. These are combined to give 36 total scenarios. \label{tab:scenTA}}
\end{table}

\subsection{Data Generation}
Data in the form of event times for activity and toxicity are generated for each simulated patient response. Following Barnett et al \cite{Barnett2024}, these event times $t_A$ and $t_T$ are generated from a bivariate log-normal model:
\[
\begin{pmatrix}
t_T \\
t_A
\end{pmatrix} \sim Lognormal_2 \left( \begin{pmatrix}
\mu_T\\
\mu_A
\end{pmatrix} , \begin{pmatrix}
\sigma^2_T & -\frac{\sigma_T \sigma_A}{2}\\
-\frac{\sigma_T \sigma_A}{2} & \sigma^2_A
\end{pmatrix} \right).
\]
with parameters $\mu_T$, $\mu_A$, $\sigma_T$, $\sigma_A$ calculated for each dose combination so that the probability of observing a toxicity event across the observation period equals that defined by the scenario specification, and similarly for activity. The pattern across cycles differs between activity and toxicity, with activity split equally across all three cycles, and 0.75 of the total probability of toxicity assigned to the first cycle. This is in line with the data generation used by Barnett et al \cite{Barnett2024}, although for simplicity here we refer only to the true probabilities of response across all cycles in the definition of scenarios. 

\subsection{Rules}
In a similar fashion to Barnett et al \cite{Barnett2024}, a set of rules reflecting those applied in such a trial are used. Let $p_{1,d_{i*j*}}$ be the probability of observing a DLT response in cycle 1 on dose $d_{i*j*}$.\\ 
Enforcement rules restrict the set of allowable doses in order to protect the safety of patients in the trial. Dose skipping restricts escalation to unexplored doses, and hard safety restricts escalation to explored doses. Stopping rules dictate the criteria for stopping the trial.\\
The parameters for the admissibility criteria for all designs are set at  $q_T=q_A=0.2$ and $\phi_T=0.3$ and $\phi_A=0.2$. \\

\subsubsection{Enforcement}
\begin{enumerate}
\item \textbf{Dose skipping:} In this dual agent case, the dose skipping rule is defined by agent. Once dose $d_{i*j*}$ has been explored, all doses $d_{ij}$ such that $i\leq i*$ an $j\leq j*+1$, or $i\leq i*+1$ and $j\leq j*$ are allowable. This ensures that dose levels of each agent are not skipped, as well as restricting escalation in both agents at once.

\item \textbf{Hard Safety}: The hard safety rule is defined by the observed number of DLT responses in the first cycle of treatment on explored doses. If there are 3 DLT responses out of 3 patients, 4 or more DLT responses out of 6 patients, or 5 or more DLT responses out of 9 patients on dose $d_{i*j*}$, then all doses $d_{ij}$ such that $i\geq i*$ and $j*\geq j$, are excluded from further exploration. This is based on an exclusion criterion of $\mathbb{P}(p_{1,d_{i*j*}}>0.3)>0.95$ using a beta-binomial model with a $Beta(1,1)$ prior.
\end{enumerate}

\subsubsection{Stopping}
\begin{enumerate}
\item \textbf{No Admissible Dose Combinations}: If no dose combinations satisfy both admissibility criteria in Section~\ref{sec:admiss}, then the trial is stopped, deeming no dose combination admissible.
\item \textbf{Lowest Dose Combination Deemed Unsafe}: If $\mathbb{P}(p_{1,d_{11}}>30\%)>0.80$ and at least one cohort of patients has been assigned to dose $d_{11}$, the trial is stopped, deeming all dose combination unsafe. \label{rule:low_unsafe}
\item \textbf{Highest Dose Combination Deemed Very Safe}: If $\mathbb{P}(p_{1,d_{IJ}}\leq 30\%)>0.80$ and at least one cohort of patients has been assigned to dose $d_{IJ}$, the trial is stopped, deeming all doses under target safety. \label{rule:hi_safe}
\item \textbf{Sufficient Information}: For a pre-defined cut-off value, $C_{suff}$, if a dose combination is recommended for the next cohort on which $C_{suff}$ patients have already been assigned in the escalation, the trial is stopped. In this case, we use $C_{suff}=30$. \label{rule:suff}
\item \textbf{Hard Safety}: If the $d_{11}$ is considered unsafe according to the hard safety enforcement rule, the trial is stopped, deeming all dose combination unsafe.
\item \textbf{Maximum Patients}: If the maximum number of patients ($n=n_{\mbox{max}}$) have been recruited, the trial is stopped. In this case, we use $n_{\mbox{max}}=60$.
\end{enumerate}

\subsection{Results}
To evaluate the merits of the three designs, their performance is measured according to criteria based both on their final recommendation of dose combination and the assignment of patients to the dose combinations. It is desirable for a design to give a high proportion of good dose combination recommendations, whilst also not exposing unnecessary levels of patients to unsafe dose combinations.

To compare the recommendations, we define certain classes of dose combinations. A dose combination is \textit{safe} if the true probability of observing a toxicity is less than or equal to 0.3 and \textit{unsafe} otherwise. A dose combination is \textit{active} if the true probability of observing an activity event is greater than or equal to 0.2 and \textit{futile} otherwise. If a dose is unsafe or futile, then it is \textit{unacceptable}. If it is both safe and active, then it is \textit{acceptable}. Within the class of acceptable dose combinations, the one dose combination with the highest true utility is defined as the \textit{correct} dose combination, however when the utility is very similar across dose combinations with the highest utility, these dose combinations are referred to as \textit{good}.

To gain an assessment of the performance, the selections are also compared to an optimal non-parametric benchmark that takes into account uncertainty of ordering \cite{Mozgunov2021a}. This is modified in this context to also account for admissibility, with further details given in the supplementary materials. 

It is worth noting that since the Joint TITE-POCRM and the Joint TITE-BLRM both use the same utility function, the definition of correct and good dose combinations are identical. However, the Joint TITE-BOIN design, here used as a model-assisted comparator, uses a different utility function and therefore in some scenarios, the definition of correct and good doses are not identical, and so direct comparisons of these metrics are not meaningful.

\begin{figure}[h!]
\centering
\begin{subfigure}{.8\textwidth}
  \centering
  \includegraphics[width=1\linewidth]{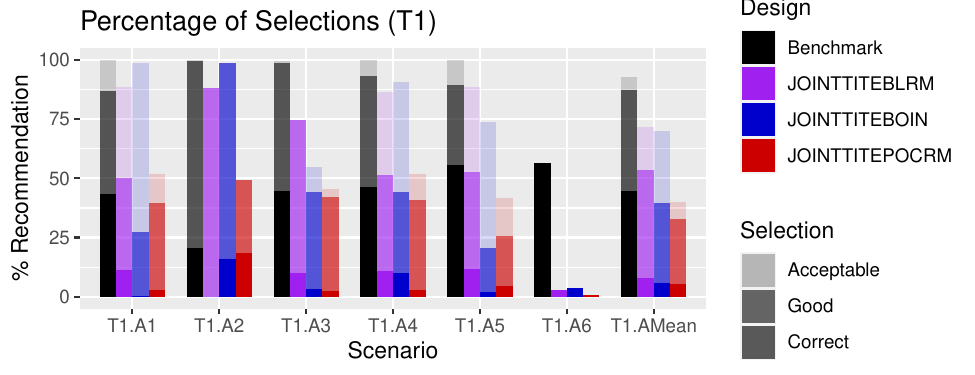}
  \caption{Percentage of Correct, Good and Acceptable selections}
  \label{fig:T1p}
\end{subfigure}\\%
\begin{subfigure}{.8\textwidth}
  \centering
  \includegraphics[width=1\linewidth]{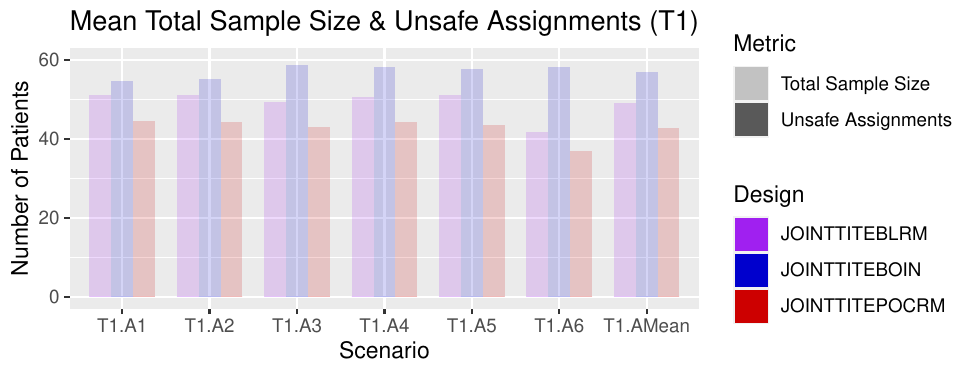}
  \caption{Mean total patients, and mean patients assigned to unsafe dose combinations.}
  \label{fig:T1s}
\end{subfigure}
\caption{Results for scenarios T1.A1 - T1.A6, including an average over these six scenarios, labelled T1.AMean.}
\label{fig:T1}
\end{figure}

\begin{figure}[h!]
\centering
\begin{subfigure}{.8\textwidth}
  \centering
  \includegraphics[width=1\linewidth]{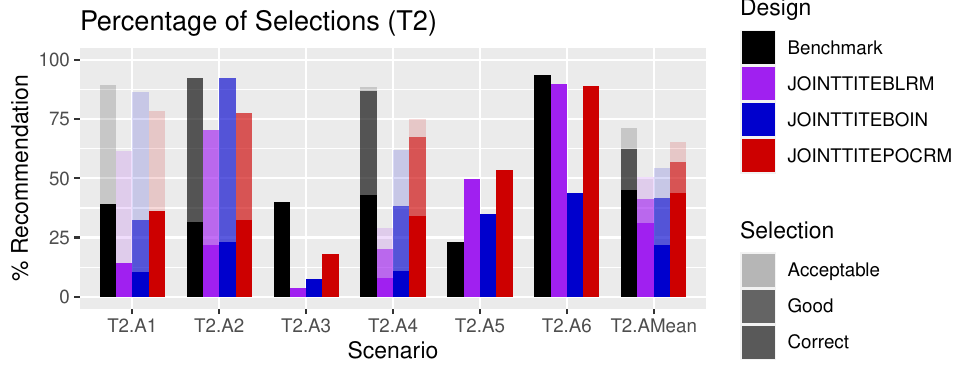}
  \caption{Percentage of Correct, Good and Acceptable selections}
  \label{fig:T2p}
\end{subfigure}\\%
\begin{subfigure}{.8\textwidth}
  \centering
  \includegraphics[width=1\linewidth]{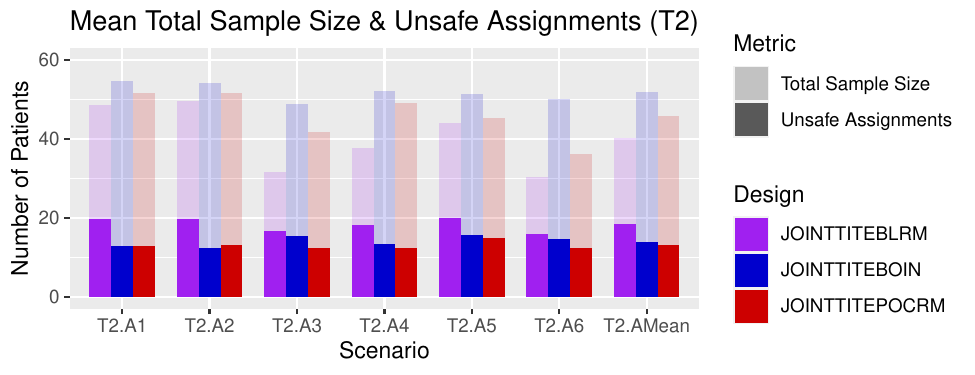}
  \caption{Mean total patients, and mean patients assigned to unsafe dose combinations.}
  \label{fig:T2s}
\end{subfigure}
\caption{Results for scenarios T2.A1 - T2.A6, including an average over these six scenarios, labelled T2.AMean.}
\label{fig:T2}
\end{figure}

\begin{figure}[h!]
\centering
\begin{subfigure}{.8\textwidth}
  \centering
  \includegraphics[width=1\linewidth]{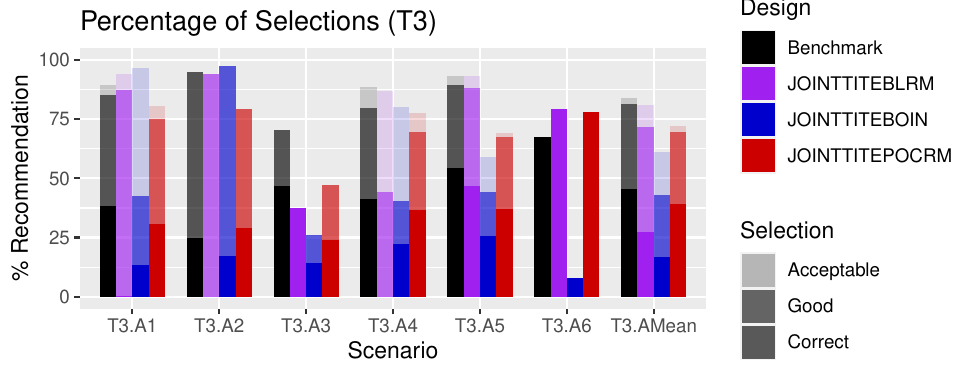}
  \caption{Percentage of Correct, Good and Acceptable selections}
  \label{fig:T3p}
\end{subfigure}\\%
\begin{subfigure}{.8\textwidth}
  \centering
  \includegraphics[width=1\linewidth]{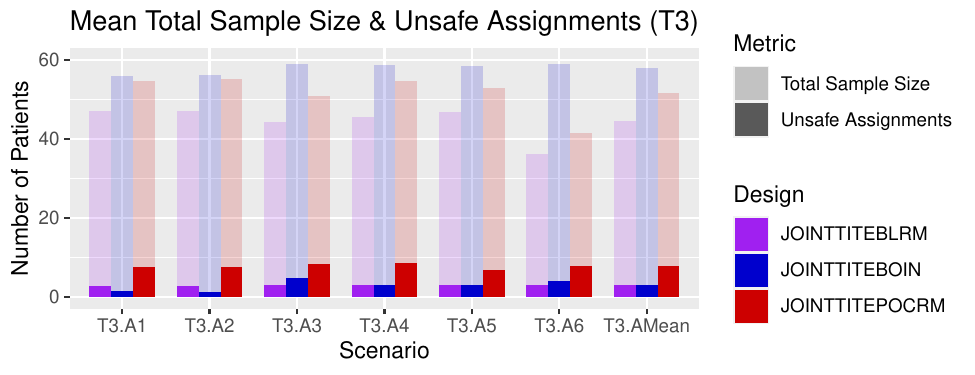}
  \caption{Mean total patients, and mean patients assigned to unsafe dose combinations.}
  \label{fig:T3s}
\end{subfigure}
\caption{Results for scenarios T3.A1 - T3.A6, including an average over these six scenarios, labelled T3.AMean.}
\label{fig:T3}
\end{figure}

\begin{figure}[h!]
\centering
\begin{subfigure}{.8\textwidth}
  \centering
  \includegraphics[width=1\linewidth]{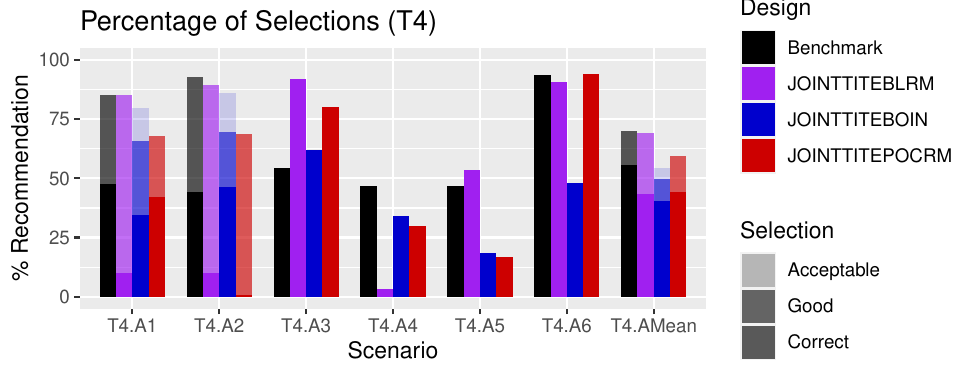}
  \caption{Percentage of Correct, Good and Acceptable selections}
  \label{fig:T4p}
\end{subfigure}\\%
\begin{subfigure}{.8\textwidth}
  \centering
  \includegraphics[width=1\linewidth]{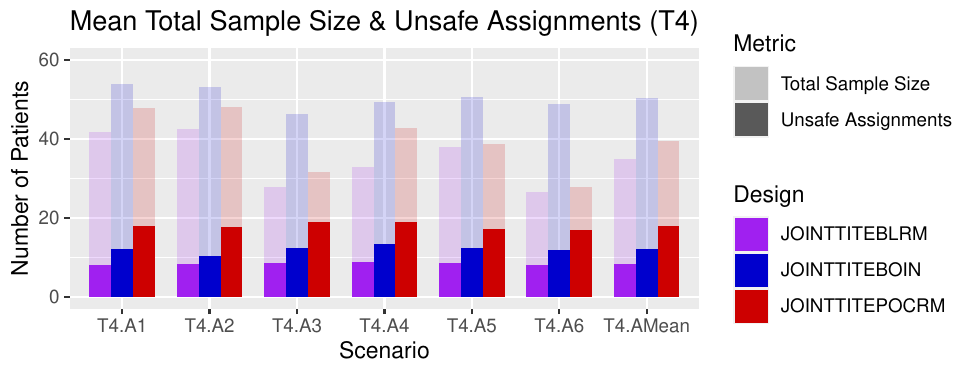}
  \caption{Mean total patients, and mean patients assigned to unsafe dose combinations.}
  \label{fig:T4s}
\end{subfigure}
\caption{Results for scenarios T4.A1 - T4.A6, including an average over these six scenarios, labelled T4.AMean.}
\label{fig:T4}
\end{figure}

\begin{figure}[h!]
\centering
\begin{subfigure}{.8\textwidth}
  \centering
  \includegraphics[width=1\linewidth]{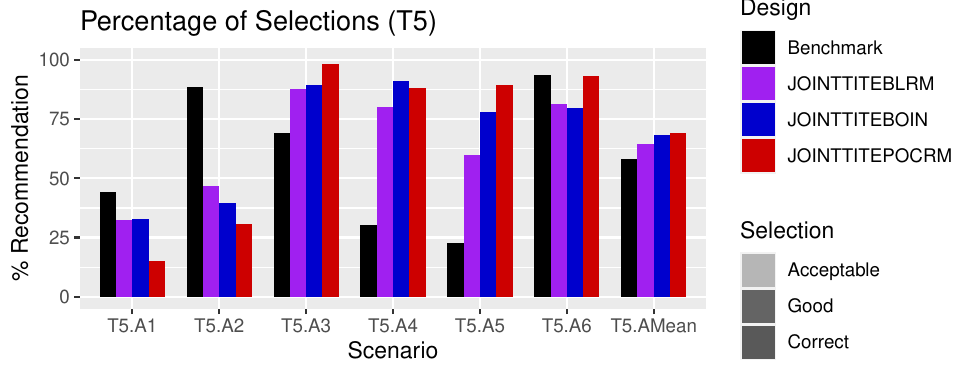}
  \caption{Percentage of Correct, Good and Acceptable selections}
  \label{fig:T5p}
\end{subfigure}\\%
\begin{subfigure}{.8\textwidth}
  \centering
  \includegraphics[width=1\linewidth]{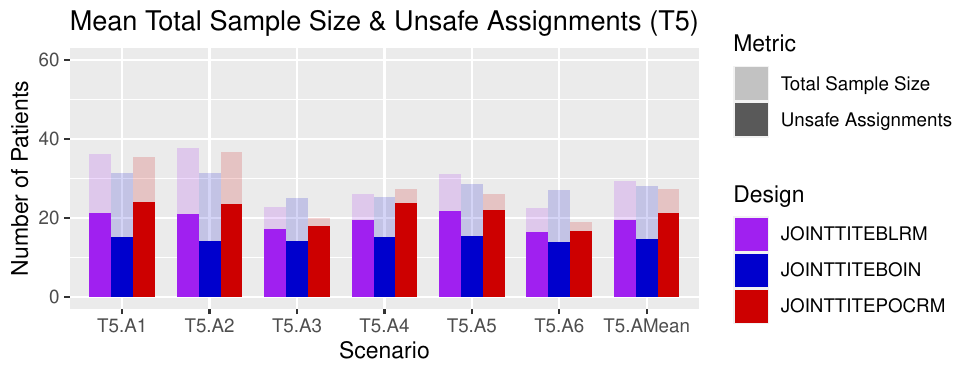}
  \caption{Mean total patients, and mean patients assigned to unsafe dose combinations.}
  \label{fig:T5s}
\end{subfigure}
\caption{Results for scenarios T5.A1 - T5.A6, including an average over these six scenarios, labelled T5.AMean.}
\label{fig:T5}
\end{figure}

\begin{figure}[h!]
\centering
\begin{subfigure}{.8\textwidth}
  \centering
  \includegraphics[width=1\linewidth]{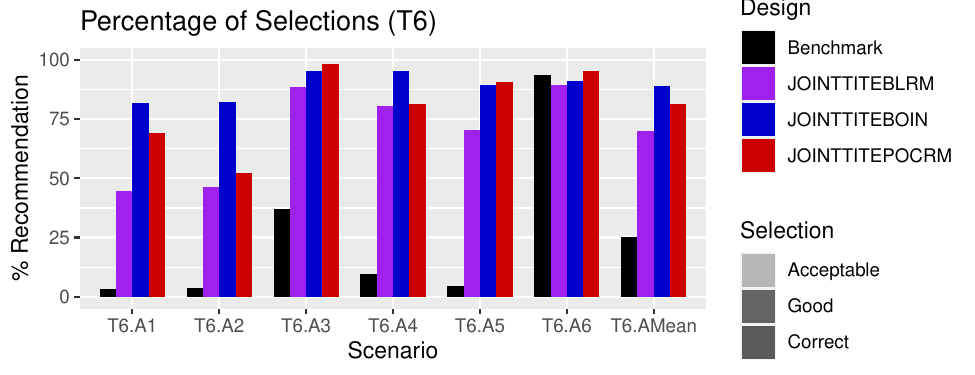}
  \caption{Percentage of Correct, Good and Acceptable selections}
  \label{fig:T6p}
\end{subfigure}\\%
\begin{subfigure}{.8\textwidth}
  \centering
  \includegraphics[width=1\linewidth]{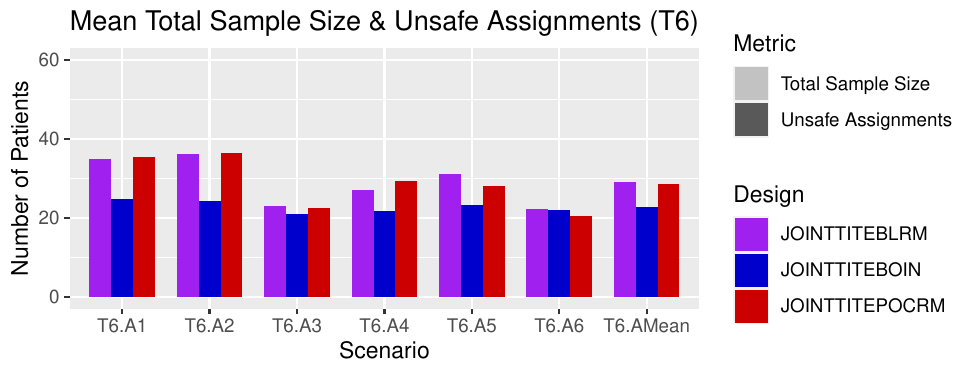}
  \caption{Mean total patients, and mean patients assigned to unsafe dose combinations.}
  \label{fig:T6s}
\end{subfigure}
\caption{Results for scenarios T6.A1 - T6.A6, including an average over these six scenarios, labelled T6.AMean.}
\label{fig:T6}
\end{figure}

\begin{figure}[h!]
\centering
\begin{subfigure}{.8\textwidth}
  \centering
  \includegraphics[width=1\linewidth]{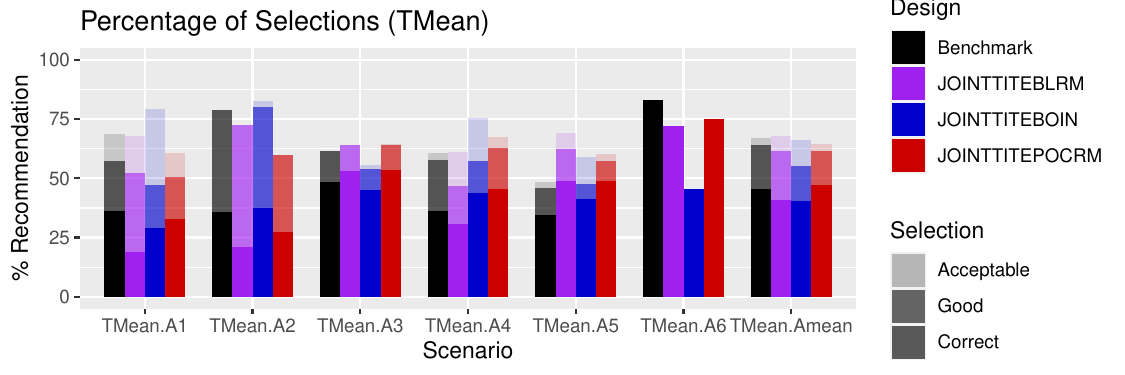}
  \caption{Percentage of Correct, Good and Acceptable selections}
  \label{fig:TMp}
\end{subfigure}\\%
\begin{subfigure}{.8\textwidth}
  \centering
  \includegraphics[width=1\linewidth]{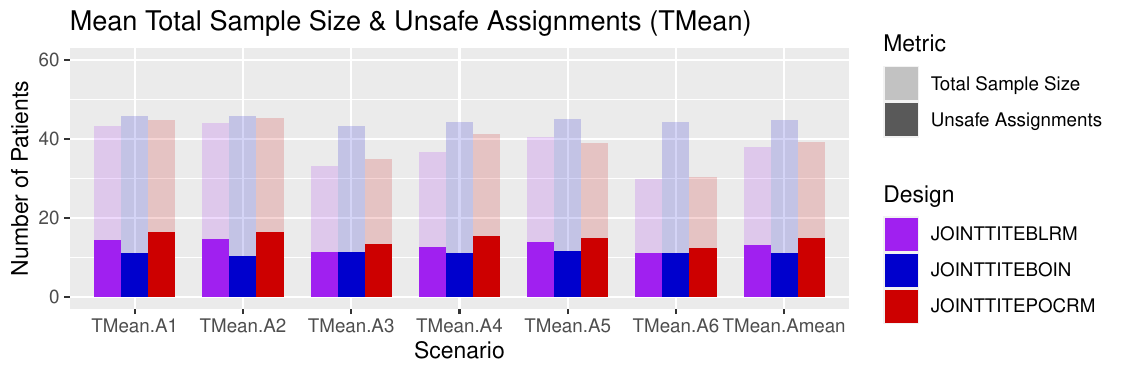}
  \caption{Mean total patients, and mean patients assigned to unsafe dose combinations.}
  \label{fig:TMs}
\end{subfigure}
\caption{Average results over toxicity scenarios, labelled TMean.A1 - TMean.A6, including an average over all 36 scenarios, labelled TMean.AMean.}
\label{fig:TM}
\end{figure}

Figures \ref{fig:T1}-\ref{fig:T6} display the results across the 36 scenarios, each figure representing one toxicity scenario, with its combination scenario with each activity scenario, and an average across these combinations. Within each figure, subfigure (a) concerns the selections, displaying the percentage of Correct, Good and Acceptable selections. Subfigure (b) concerns the assignments, displaying the average total sample size and the average number of patients assigned to unsafe doses. Figure \ref{fig:TM} displays the mean of these results over the toxicity scenarios for each activity scenario, and an average over all 36 scenarios. Full details of the selections and assignments are available in the supplementary materials.

Since the overall average displayed in the final columns of Figures \ref{fig:TMp} and \ref{fig:TMs} show only a small difference across the three approaches, one might assume that there is comparable performance. However, there is a large amount of variation when one looks in more detail at the individual scenarios.

For example, in scenarios T1.A1, T1.A2, T1.A3, T1.A4 and T1.A5, the Joint TITE-BLRM considerably outperforms the Joint TITE-POCRM. In these scenarios, all dose combinations are safe, and the correct dose combination is at the highest level of Agent $W_2$, with good dose combinations surrounding it. In these scenarios, the somewhat poor performance of  the Joint TITE POCRM is due to it incorrectly stopping early due to stopping rule \ref{rule:low_unsafe}, that the highest dose is deemed too safe. The Joint TITE-BOIN shows a performance between the two model-based designs in terms of selections, but with a larger sample size.

In contrast, it can be seen in Figure \ref{fig:T2p} in scenario T2.A1 and T2.A4, the Joint TITE-POCRM considerably outperforms the Joint TITE-BLRM. In both of these scenarios, dose combination $d_{14}$ is the OBD. In T2.A4, the Joint TITE-BLRM incorrectly concludes that there is no admissible dose in 57\% of simulations, despite a number of dose combinations being truly active and safe. The Joint TITE-POCRM however, shows a better performance. It is also apparent from Figure \ref{fig:T2s} that the Joint TITE-BLRM assigns more patients on average to unsafe dose combinations, despite a smaller average overall sample size. As a comparator, the Joint TITE-BOIN here shows a poor performance when there are truly no admissible dose combinations, being more willing to recommend an inactive dose combination.

Figure \ref{fig:T3p} shows some interesting contrasting behaviour of the two model-based approaches. In T3.A3 and T3.A5, the Joint TITE-BLRM shows superior performance in terms of correct selections, however in T3.A1  T3.A2 and T3.A4, the Joint TITE-POCRM shows superior performance in terms of correct selections, and the Joint TITE-BLRM fails to select the correct dose combination at all. Here the correct dose combination is on the lowest dose of Agent $W_1$, and the Joint TITE-BLRM seems to favour good doses on the higher dose level of Agent $W_1$. The Joint TITE-BOIN again is displaying behaviour whereby it is more willing to recommend inactive dose combinations.

Figure \ref{fig:T4p} again shows the contrast in the behaviour. The average performance across T4 is very similar in terms of correct selections, but the Joint TITE-BLRM has more good selections. However. for example T4.A4 and T4.A5 show very different performances. In T4.A4, the Joint TITE-BLRM shows a very poor performance. The only acceptable/good/correct dose is $d_{12}$, and the Joint TITE-BLRM only selects this dose combination in 3\% of simulations, selecting the inactive $d_{21}$ in 18\% of simulations and no admissible dose in 74\% of simulations. Whereas the Joint TITE-POCRM selects the correct dose combination in 30\% of simulations. In contrast, in scenario T4.A5, dose $d_{21}$ is the only acceptable/good/correct dose, and the Joint TITE-POCRM only selects this dose in 17\% of simulations, selecting the inactive $d_{12}$ in 5\% of simulations and no admissible dose in 69\% of simulations. The Joint TITE-BLRM however recommends the correct dose in 54\% of simulations. This comparison is of particular interest, as from the starting dose of $d_{22}$ being unsafe, these two scenarios illustrate the differences when it is the increase in Agent $W_1$ or Agent $W_2$ that is driving the toxicity. The Joint TITE-POCRM shows a more balanced performance, whereas the Joint TITE-BLRM shows a clear preference for Agent $W_1$.

Since in T5, only the lowest dose combination is safe, only in scenarios T5.A1 and T5.A2 are we seeking to find the correct dose combination, and in the other four T5 scenarios we seek to declare no admissible doses. In Figure \ref{fig:T5}, the Joint TITE-POCRM shows superior performance in T5.A5 and T5.A6, where the Joint TITE-BLRM is more likely to recommend the inactive dose combinations than the Joint TITE-POCRM. The Joint TITE-BLRM identifies the lowest dose combination as the OBD in scenarios T5.A1 and T5.A2 more than the Joint TITE-POCRM.

In the T6 scenarios, all dose combinations are unsafe, and therefore the correct recommendation is either stopping for no admissible dose, or for a safety stopping rule. Here, Figure~\ref{fig:T6} illustrates the performance of the methods in these scenarios. It is clear to see that the Joint TITE-BOIN design not only correctly stops in more simulations, but also stops sooner, with much smaller mean sample sizes across the scenarios. The Joint TITE-POCRM and Joint TITE-BLRM show similar performances to each in T6.A2, T6.A4 and T6.A6, however the across the rest of the scenarios the Joint TITE-POCRM shows superior performance. Both model-based methods show poore performances in the scenarios where the lowest dose combinations are active.

Figure \ref{fig:TM} shows the average results over toxicity scenarios, including an average over all 36 scenarios. Whilst conclusions cannot be drawn in cases where the averages are similar, it is of interest to note those activity scenarios where the results are contrasting. Of note is A4, where the Joint TITE-POCRM shows superior performance. Here, in three out of the six individual scenarios, the OBD was at the the 600mg dose of Agent $W_1$, which the Joint TITE-BLRM finds harder to identify. The Joint TITE-BOIN shows inferior performance and increased sample size in A6, where all but the highest dose combination is inactive. This is driven by the reluctance to stop for inactivity by the Joint TITE-BOIN design.

\subsection{Changing the size of the dosing grid}
So far we have considered scenarios whereby the dosing grid has dimensions $2 \times 5$, Agent $W_1$ having two dose levels, 1200mg and a fall-back of 600mg, and Agent $W_2$ has 5 dose levels to explore. Now we extend the set of scenarios to include those where the dosing grid has dimensionality $3 \times 3$. Here we extend the range of Agent $W_1$ to include a higher dose of 1800mg, and restrict the range of Agent $W_2$ to the lower three doses.

A smaller range of scenarios are considered in this setting, labelled S1-S6, with their definitions given in Table~\ref{tab:3by3_scens}. These scenarios are chosen to give a range of relationships across the grid, and to further explore the results seen in the $2 \times 5$ grid.

The starting dose remains $d_{22}$, and the prior distribution require little amendments, only the POCRM requires the final value in the skeleton to be removed. The partial orders considered in the POCRM follow the same pattern as recommended by Wages et al. \cite{Wages2013b}, detailed in the supplementary materials.

\begin{table}[ht!]
{\footnotesize
\begin{tabular}{lllllllllllllll}
                                        &                                                                                               & \multicolumn{3}{c}{Toxicity}         &  &                                                                                               & \multicolumn{3}{c}{Activity}          &  &                                                                                               & \multicolumn{3}{c}{Utility}                                                          \\ \cline{2-5} \cline{7-10} \cline{12-15} 
                                        & \multicolumn{1}{c|}{\multirow{2}{*}{\begin{tabular}[c]{@{}c@{}}Agent $W_1$ \\ (mg)\end{tabular}}} & \multicolumn{3}{c}{Agent $W_2$ (kBq/kg)} &  & \multicolumn{1}{c|}{\multirow{2}{*}{\begin{tabular}[c]{@{}c@{}}Agent $W_1$ \\ (mg)\end{tabular}}} & \multicolumn{3}{c}{Agent $W_2$ (kBq/kg)} &  & \multicolumn{1}{c|}{\multirow{2}{*}{\begin{tabular}[c]{@{}c@{}}Agent $W_1$ \\ (mg)\end{tabular}}} & \multicolumn{3}{c}{Agent $W_2$ (kBq/kg)}                                                 \\ \cline{3-5} \cline{8-10} \cline{13-15} 
                                        & \multicolumn{1}{c|}{}                                                                         & 50          & 75         & 100       &  & \multicolumn{1}{c|}{}                                                                         & 50         & 75         & 100        &  & \multicolumn{1}{c|}{}                                                                         & 50                   & 75                            & 100                           \\ \cline{2-5} \cline{7-10} \cline{12-15} 
\multicolumn{1}{c}{\multirow{3}{*}{S1}} & \multicolumn{1}{l|}{600}                                                                      & 0.1         & 0.15       & 0.2       &  & \multicolumn{1}{l|}{600}                                                                      & 0.05       & 0.1        & 0.15       &  & \multicolumn{1}{l|}{600}                                                                      & 0.017                & 0.0505                        & 0.084                         \\
\multicolumn{1}{c}{}                    & \multicolumn{1}{l|}{1200}                                                                     & 0.15        & 0.2        & 0.3       &  & \multicolumn{1}{l|}{1200}                                                                     & 0.2        & 0.25       & 0.3        &  & \multicolumn{1}{l|}{1200}                                                                     & \textit{0.1505}      & \textit{0.184}                & {\ul \textit{0.201}}          \\
\multicolumn{1}{c}{}                    & \multicolumn{1}{l|}{1800}                                                                     & 0.2         & 0.3        & 0.45      &  & \multicolumn{1}{l|}{1800}                                                                     & 0.35       & 0.4        & 0.45       &  & \multicolumn{1}{l|}{1800}                                                                     & {\ul \textit{0.284}} & {\ul \textit{\textbf{0.301}}} & -0.7885                       \\ \cline{2-5} \cline{7-10} \cline{12-15} 
\multirow{3}{*}{S2}                     & \multicolumn{1}{l|}{600}                                                                      & 0.05        & 0.09       & 0.11      &  & \multicolumn{1}{l|}{600}                                                                      & 0.05       & 0.08       & 0.10       &  & \multicolumn{1}{l|}{600}                                                                      & 0.034                & 0.050                         & 0.064                         \\
                                        & \multicolumn{1}{l|}{1200}                                                                     & 0.07        & 0.13       & 0.25      &  & \multicolumn{1}{l|}{1200}                                                                     & 0.10       & 0.12       & 0.15       &  & \multicolumn{1}{l|}{1200}                                                                     & 0.077                & 0.077                         & 0.068                         \\
                                        & \multicolumn{1}{l|}{1800}                                                                     & 0.15        & 0.20       & 0.30      &  & \multicolumn{1}{l|}{1800}                                                                     & 0.20       & 0.30       & 0.40       &  & \multicolumn{1}{l|}{1800}                                                                     & \textit{0.151}       & {\ul \textit{0.234}}          & {\ul \textit{\textbf{0.301}}} \\ \cline{2-5} \cline{7-10} \cline{12-15} 
\multirow{3}{*}{S3}                     & \multicolumn{1}{l|}{600}                                                                      & 0.05        & 0.10       & 0.15      &  & \multicolumn{1}{l|}{600}                                                                      & 0.10       & 0.20       & 0.30       &  & \multicolumn{1}{l|}{600}                                                                      & 0.084                & \textit{0.167}                & {\ul \textit{0.251}}          \\
                                        & \multicolumn{1}{l|}{1200}                                                                     & 0.20        & 0.25       & 0.30      &  & \multicolumn{1}{l|}{1200}                                                                     & 0.20       & 0.30       & 0.40       &  & \multicolumn{1}{l|}{1200}                                                                     & \textit{0.134}       & {\ul \textit{0.218}}          & {\ul \textit{\textbf{0.301}}} \\
                                        & \multicolumn{1}{l|}{1800}                                                                     & 0.40        & 0.45       & 0.50      &  & \multicolumn{1}{l|}{1800}                                                                     & 0.30       & 0.40       & 0.50       &  & \multicolumn{1}{l|}{1800}                                                                     & -0.922               & -0.839                        & -0.755                        \\ \cline{2-5} \cline{7-10} \cline{12-15} 
\multirow{3}{*}{S4}                     & \multicolumn{1}{l|}{600}                                                                      & 0.20        & 0.25       & 0.30      &  & \multicolumn{1}{l|}{600}                                                                      & 0.20       & 0.25       & 0.45       &  & \multicolumn{1}{l|}{600}                                                                      & \textit{0.134}       & \textit{0.168}                & {\ul \textit{\textbf{0.351}}} \\
                                        & \multicolumn{1}{l|}{1200}                                                                     & 0.40        & 0.45       & 0.50      &  & \multicolumn{1}{l|}{1200}                                                                     & 0.30       & 0.40       & 0.50       &  & \multicolumn{1}{l|}{1200}                                                                     & -0.922               & -0.839                        & -0.755                        \\
                                        & \multicolumn{1}{l|}{1800}                                                                     & 0.50        & 0.55       & 0.60      &  & \multicolumn{1}{l|}{1800}                                                                     & 0.35       & 0.50       & 0.60       &  & \multicolumn{1}{l|}{1800}                                                                     & -0.905               & -0.772                        & -0.688                        \\ \cline{2-5} \cline{7-10} \cline{12-15} 
\multirow{3}{*}{S5}                     & \multicolumn{1}{l|}{600}                                                                      & 0.05        & 0.10       & 0.20      &  & \multicolumn{1}{l|}{600}                                                                      & 0.05       & 0.15       & 0.30       &  & \multicolumn{1}{l|}{600}                                                                      & 0.034                & 0.117                         & {\ul \textit{\textbf{0.234}}} \\
                                        & \multicolumn{1}{l|}{1200}                                                                     & 0.15        & 0.25       & 0.35      &  & \multicolumn{1}{l|}{1200}                                                                     & 0.10       & 0.25       & 0.40       &  & \multicolumn{1}{l|}{1200}                                                                     & 0.051                & {\ul \textit{0.168}}          & -0.806                        \\
                                        & \multicolumn{1}{l|}{1800}                                                                     & 0.30        & 0.40       & 0.45      &  & \multicolumn{1}{l|}{1800}                                                                     & 0.20       & 0.35       & 0.45       &  & \multicolumn{1}{l|}{1800}                                                                     & \textit{0.101}       & -0.872                        & -0.789                        \\ \cline{2-5} \cline{7-10} \cline{12-15} 
\multirow{3}{*}{S6}                     & \multicolumn{1}{l|}{600}                                                                      & 0.05        & 0.20       & 0.40      &  & \multicolumn{1}{l|}{600}                                                                      & 0.05       & 0.20       & 0.35       &  & \multicolumn{1}{l|}{600}                                                                      & 0.034                & {\ul \textit{0.134}}          & -0.872                        \\
                                        & \multicolumn{1}{l|}{1200}                                                                     & 0.10        & 0.25       & 0.45      &  & \multicolumn{1}{l|}{1200}                                                                     & 0.10       & 0.25       & 0.40       &  & \multicolumn{1}{l|}{1200}                                                                     & 0.067                & {\ul \textit{0.168}}          & -0.839                        \\
                                        & \multicolumn{1}{l|}{1800}                                                                     & 0.15        & 0.30       & 0.50      &  & \multicolumn{1}{l|}{1800}                                                                     & 0.15       & 0.30       & 0.45       &  & \multicolumn{1}{l|}{1800}                                                                     & 0.101                & {\ul \textit{\textbf{0.201}}} & -0.805                        \\ \cline{2-5} \cline{7-10} \cline{12-15} 
\end{tabular}}
\caption{Toxicity and Activity probabilities and Utility values for scenarios S1-S6. Acceptable dose combinations are highlighted in \textit{italics}, good dose combinations are \uu{underlined} and correct dose combinations are highlighted in \textbf{boldface}. \label{tab:3by3_scens}}
\end{table}

\begin{figure}[h!]
\centering
\begin{subfigure}{.8\textwidth}
  \centering
  \includegraphics[width=1\linewidth]{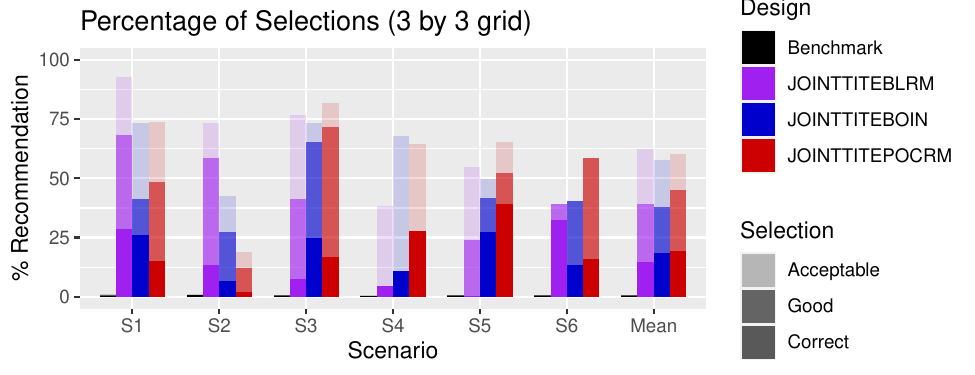}
  \caption{Percentage of Correct, Good and Acceptable selections}
  \label{fig:33p}
\end{subfigure}\\%
\begin{subfigure}{.8\textwidth}
  \centering
  \includegraphics[width=1\linewidth]{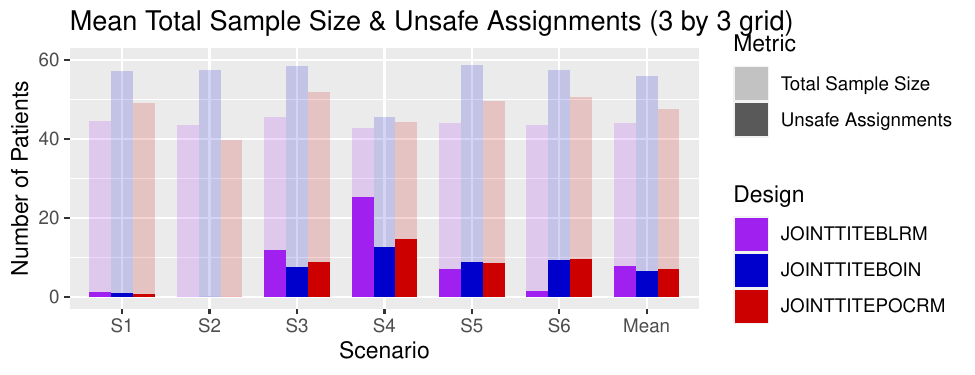}
  \caption{Mean total patients, and mean patients assigned to unsafe dose combinations.}
  \label{fig:33s}
\end{subfigure}
\caption{Results for the scenarios with 3 by 3 grids, including an average over these scenarios.}
\label{fig:33}
\end{figure}

The results for the simulation study of $3 \times 3$ grids are shown in Figure \ref{fig:33}. There are noticeable differences between the Joint TITE-BLRM and Joint TITE-POCRM across this range of scenarios. In S1, the Joint TITE-BLRM recommends the correct dose in 29\% of the simulations, compared to only 15\% for the Joint TITE-POCRM. The Joint TITE-BLRM also has a higher percentage of good and acceptable selections.

In S2, again the Joint TITE-BLRM shows more promising behaviour, recommending a good dose combination in 59\% of simulations compared to the 12\% of the Joint TITE-POCRM. This stark difference is due to the POCRM incorrectly stopping the trial for no admissible doses or the highest dose being too safe in 74\% of simulations. Here only the 1800mg dose of Agent $W_1$ is active, and all doses are safe. The behaviour displayed is similar to that seen in scenario T1.A5, that the Joint TITE-POCRM struggles to identify good doses when only the highest dose level of Agent $W_1$ is active.

In S3 - S6 however, the reverse is true, and the Joint TITE-POCRM show superior performance to the Joint TITE-BLRM. In S3, where the 1800mg dose of Agent $W_1$ is unsafe, and only the lowest dose combination is inactive, the Joint TITE-POCRM recommends a good dose in 72\% of simulations, compared to 41\% for the Joint TITE-BLRM. Interestingly, the Joint TITE-POCRM favours recommending $d_{13}$, whereas the Joint TITE-BLRM favours recommending $d_{21}$ and $d_{22}$. In S4, where only the 600mg dose of Agent $W_1$ is safe, the BLRM incorrectly recommends the unsafe $d_{21}$ in 38\% of simulations, and also favours in recommendation the dose combination $d_{11}$ only recommending the correct $d_{13}$ in 5\% of simulations, hence a poorer performance than the POCRM, which favours in recommendation the dose combination $d_{12}$ at 30\%, followed by the correct $d_{13}$ in 28\% of simulations.

In S5, the acceptable dose combinations lie along the off-diagonal of the grid, increasing utility as the dose of Agent $W_2$ increases. The Joint TITE-POCRM recommends the correct dose in 39\% of simulations, whereas the Joint TITE-BLRM only recommends this dose in 1\% of simulations, giving more favour to $d_{31}$ and $d_{22}$, and has over twice as many simulations recommending no admissible dose than the Joint TITE-POCRM, at 28\%. In S6, the highest dose of Agent $W_2$ in unsafe and the lowest dose of Agent $W_2$ is inactive, leaving only the middle dose. Here, the Joint TITE-BLRM recommend the correct dose of $d_{32}$ in more simulations than the Joint TITE-POCRM (33\% vs 16\%), the POCRM recommends a good dose in 59\% of simulations compared to the BLRM's 39\%. The downfall of the Joint TITE-BLRM in this scenario is in its over-recommendation of $d_{31}$, which is inactive.

The Joint TITE-BOIN again shows a larger sample size than the model-based designs, fewer acceptable selections than the leading model-based design in each scenario, but in general quite balanced results across the scenarios.

\section{Discussion} \label{sec:dis}
In this work, two model-based approaches for a Phase I/II dose-finding trial with dual agents and late-onset endpoints are presented. One is based on the BLRM, and the other the POCRM. Both approaches show promise, but with contrasting behaviours in a subset of considered scenarios.

These contrasting behaviours are inherent due to the differences in nature of the two designs, accentuated by the dual time-to-event endpoint. Since the Joint TITE-POCRM has a fixed skeleton that can be mapped onto the dosing grid for the different partial orderings, this means that there is no fixed prior for each dose combination. In contrast, the Joint TITE-BLRM must by definition have a fixed prior for each dose combination. However, the Joint TITE-BLRM models interaction between the two agents and interaction between the two endpoints, whilst the Joint TITE-POCRM only models the interaction between the two endpoints in its second stage. Therefore both approaches offer flexibility in different ways, and this is illustrated in the simulation results presented here. In the case where there are two-dose levels of $W_1$, the Joint TITE-BLRM is better at identifying when the fall-back dose of Agent $W_1$ is inactive, and in general explores the higher dose of Agent $W_1$ well, only deviating to the lower dose if the higher dose is deemed unsafe. The POCRM on the other hand, explores the grid more, and will explore the lower dose of Agent $W_1$ more often.

One obvious limitation of both methods is their computational intensity. The complex models require MCMC methods to compute posterior distributions for parameters. The large number of parameters on which priors must be elicited is especially problematic for the BLRM as the high dimensionality of the prior calibration greatly increases the computational load. The reliance on the prior calibration should also be noted. Here, both methods were subject to the same method of calibration an hence both given the fair chance to perform. Within the results of this calibration, it was clear to see that the results were particularly sensitive to some changes in hyper-parameter values and robust to others. Therefore the prior specification must be considered carefully in the context of the trial.

The models themselves also rely on a number of assumptions. For example, the monotonicity of both the activity and toxicity with dose. It may be of interest for example to consider non-monotonicity of activity. Whilst the framework of both approaches could be adapted to accommodate non-monotonicity, the way in which this should be done would depend upon any further assumptions on the dose-response relationship.

Whilst we endeavour to present a comprehensive study of the proposed methods, there are of course other extensions that could be considered. Sensitivity to the distribution of time-to-event, number of cycles, cohort size and sample size could all be considered of interest and would be especially pertinent in application to a trial.

\section*{Acknowledgements}
Funding was received from UK Medical Research Council (MC\_UU\_00040/03). For the purpose of open access, the author has applied a Creative Commons Attribution (CC BY) licence to any Author Accepted Manuscript version arising.

\section*{Data availability}
Software in the form of R code used to produce the presented results is available at \url{https://github.com/helenyb/Dual_Agent_Joint_TITE_CRM}

\end{document}


\maketitle

\section{TITE-BOIN12}
The original proposal of TITE-BOIN12 \cite{Zhou2022} is for single agent trials, here we have extended the approach in line with the dual agent BOIN design \cite{lin2017}. As a model-assisted design,  a model is used at each dose level to make escalation decisions, rather than modelling the dose-response relationship like the two mode-based designs.

In order to calculate the utility criterion at a dose combination, a utility value is given to each of the four possible outcomes of activity and toxicity, $\Psi_{Y^{(T)} Y^{(A)}}$, for $(Y^{(T)} , Y^{(A)} ) \in \{ (0,1),(0,0), (1,0), (1,1)\}$.

Here we retain the notation of $w_{\ell}^{(A)}$ and $w_{\ell}^{(T)}$, and introduce the variable $\delta_{\ell}^{(T)}$, equalling 1 if $Y_{\ell}^{(T)} $ has been ascertained (either an outcome observed, or patient $\ell$ has reached the end of their follow-up period), and 0 otherwise. The variable $\delta_{\ell}^{(A)}$ is defined in the same way for activity.

An approximated likelihood is then used at each dose level, using $x$, the number of quasi-events:
\begin{align*}
x = &\frac{1}{100} \sum_{a=0}^{1} \sum_{b=0}^1 \left\{ \Psi_{ab} \sum_{\ell=1}^{n} \mathbb{I}(Y_{\ell}^{(T)}  =a) \mathbb{I}(Y_{\ell}^{(A)}  =b) \delta_{\ell}^{(T)} \delta_{\ell}^{(A)} \right. \\ 
&+ \Psi_{ab} \sum_{\ell=1}^{n} \mathbb{I}(Y_{\ell}^{(T)}  =a) P(Y_{\ell}^{(A)}  =b|\delta_{\ell}^{(A)} =0) \delta_{\ell}^{(T)} (1-\delta_{\ell}^{(A)}) \\
&+ \Psi_{ab} \sum_{\ell=1}^{n} P(Y_{\ell}^{(T)}  =a |\delta_{\ell}^{(T)}=0) \mathbb{I}(Y_{\ell}^{(A)}  =b) (1-\delta_{\ell}^{(T)} )\delta_{\ell}^{(A)}\\
 &+ \left.  \Psi_{ab} \sum_{\ell=1}^{n} P(Y_{\ell}^{(T)}  =a |\delta_{\ell}^{(T)}=0) P(Y_{\ell}^{(A)}  =b|\delta_{\ell}^{(A)} =0)(1-\delta_{\ell}^{(T)} )(1-\delta_{\ell}^{(A)}) \right\}.
\end{align*}

Here, $P(Y_{\ell}^{(T)} =1 |\delta_{\ell}^{(T)}=0)=\frac{\hat{\pi}_T(1-w_{\ell}^{(T)})}{1-\hat{\pi}_T w_{\ell}^{(T)}}$ where $\hat{\pi}_T=\frac{\sum_{\ell=1}^{n}\delta_{\ell}^{(T)} y_{\ell}^{(T)}}{\sum_{\ell=1}^{n}\delta_{\ell}^{(T)} y_{\ell}^{(T)}+\sum_{\ell=1}^{n}\delta_{\ell}^{(T)}(1- y_{\ell}^{(T)})+\sum_{\ell=1}^{n}w_{\ell}^{(T)}(1-\delta_{\ell}^{(T)})}$ is the current estimate of $\pi_T$. Similarly for activity, replacing $T$ with $A$.

The posterior distribution of the utility ($u^*$) at each dose is:
\[
u^* |D \sim Beta(1+x , 1+n-x)
\]
Since it is assumed $x \sim Bin(n, u^*)$ with prior $u^* \sim Beta(1,1)$.

The utility criterion measure used in the dosing decision is $P(u^*>u_b|D)$, with the objective to choose the dose out of the defined set that maximises the probability of being above some utility threshold, $u_b=\underline{u}+(100-\underline{u})/2$, where $\underline{u}=\Psi_{01}\phi_A(1-\phi_T) +\Psi_{00}*(1-\phi_T)(1-\phi_A)+ \Psi_{11}\phi_{T}\phi_A$.

The de-escalation set $\mathcal{A}_D$, stay set $\mathcal{A}_S$ and escalation set $\mathcal{A}_E$ for dose combination $d_{ij}$ are defined as:
\[
\mathcal{A}_D = \{ d_{i-1,j}, d_{i,j-1} \}
\]
\[
\mathcal{A}_S = \{ d_{i-1,j}, d_{i,j-1} , d_{ij} \}
\]
\[
\mathcal{A}_E = \{ d_{i-1,j}, d_{i,j-1}, d_{i+1,j-1}, d_{i-1,j+1} ,d_{ij}, d_{i+1,j}, d_{i,j+1} \}
\]
These are defined so as to bring together the methodology of the TITE-BOIN12 design for single agents, and the dual agent BOIN design. 

The choice of the next dose combination is then made at each dosing decision point in the following way. Consider the current dose as $d_{ij}$. 
\begin{itemize}
\item If $\hat{\pi_T}(d_{ij}) \geq \lambda_d$, then the decision is made to de-escalate to the dose combination in $\mathcal{A}_D$ that is admissible and that maximises $P(u^*>u_b|D)$.
\item If $ \lambda_e < \hat{\pi_T}(d_{ij})  < \lambda_d$, then choose the dose combination in the set $\mathcal{A}_S$ that is admissible and that maximises $P(u^*>u_b|D)$.
\item Otherwise, choose the dose combination in the set $\mathcal{A}_E$ that is admissible and that maximises $P(u^*>u_b|D)$.
\end{itemize}

\pagebreak
\section{Prior Distributions} \label{sec:prior}
Since all methods are Bayesian, we must specify prior distributions for the parameters of the models. 

\subsubsection{Prior Calibration} \label{sec:prior_cal}
For both the Joint TITE-POCRM and the Joint TITE-BLRM, some of the hyper-parameters that specify the prior distribution are calibrated (see for example \cite{Mozgunov2021}) to give overall good performance in terms of Percentage of Correct Selections (PCS). The process of the calibration is as follows:
\begin{enumerate}
\item Consider only the safety model, and calibrate the safety part of the prior via a grid search over a number of plausible safety scenarios, choosing the combination of values giving the highest geometric mean PCS. 
\item Consider the full joint model of safety and activity, fixing the prior for safety as chosen previously and calibrate the activity part of the prior, considering a number of reasonable settings over a number of safety/activity scenarios, choosing the combination of values giving the highest geometric mean PCS. \label{num:act_cal}
\item Consider the full joint model of safety and activity, fixing the prior for activity as chosen previously and calibrate the safety part of the prior, considering a number of reasonable settings over a number of safety/activity scenarios, choosing the combination of values giving the highest geometric mean PCS. \label{num:tox_cal}
\item Repeat steps \ref{num:act_cal} and \ref{num:tox_cal} until the improvement in PCS is below a pre-specified bound.
\end{enumerate}

\subsubsection{Joint TITE-POCRM}
The skeletons for activity and toxicity are as follows\\ $x^{(A)}=(0.300, 0.402, 0.501, 0.593, 0.673, 0.741, 0.797, 0.842, 0.878, 0.906)
$ and \\$x^{(T)}=(0.261, 0.300, 0.340, 0.381, 0.422, 0.462, 0.501, 0.538, 0.574, 0.609)$

The priors used in Part 1, for the independent one-parameter models are in-line with the original POCRM \cite{wages2011}, $\alpha^{(A)} \sim N(0, 1.34^2)$ and $\alpha^{(T)} \sim N(0, 1.34^2)$ 
The following prior is used in Part 2, in-line with the Joint-TITE-CRM \cite{Barnett2024}:
\begin{equation}\label{eq:mvn}
\begin{pmatrix}
\beta_0 \\
\log (\beta_1)
\end{pmatrix} \sim N_2 \left( \begin{pmatrix}
c_1\\
c_2
\end{pmatrix} , 
\begin{pmatrix}
v_1 & 0 \\
0 & v_2 
\end{pmatrix} \right).
\end{equation} 
for both toxicity and activity. Calibrated via the procedure outline in Section~\ref{sec:prior_cal}, the values of hyper-parameters used are given in Table~\ref{tab:hyp}. We used a vague prior for $\psi \sim N(0,100)$.

\begin{table}[h!]
\begin{tabular}{lllllll}
                    &                &        & $c_1$ & $c_2$ & $v_1$ & $v_2$ \\ \hline
               \multirow{2}{*}{POCRM} &&      Toxicity &  -4  &    2.5  &     1 & 0.25 \\ \cline{3-7} 
            &&   Activity & -4    &  2    &  1.5   &  0.5  \\ \hline
                    
\multirow{4}{*}{BLRM} & \multirow{2}{*}{Agent $W_1$} &Toxicity & -6    & 0     & 1     & 0.5   \\ \cline{3-7}   
                      &                          & Activity & -3    & -5    & 1     & 4     \\ \cline{2-7} 
                      & \multirow{2}{*}{Agent $W_2$} & Toxicity &-5.5  & 1     & 1     & 0.5   \\ \cline{3-7}  
                      &                          &  Activity &  -4    & -5    & 1     & 4   \\ \hline
\end{tabular}
\caption{Values of hyperparameter for the Joint TITE-POCRM and Joint TITE-BLRM. \label{tab:hyp}}
\end{table}

As well as prior distributions, the Joint TITE-POCRM also requires the specification of partial orderings for both activity and toxicity. The set of considered partial orderings is the same for both activity and toxicity. Here, the specifications of partial orderings follow those of Wages et al. \cite{Wages2013b}.
For the $2 \times 5$ grid:\\
\textbf{Order 1 (Rows):}
\[
R(d_{11}) \leq 
R(d_{12}) \leq 
R(d_{13}) \leq 
R(d_{14}) \leq 
R(d_{15}) \leq 
R(d_{21}) \leq 
R(d_{22}) \leq 
R(d_{23}) \leq
R(d_{24}) \leq 
R(d_{25}) 
\]
\textbf{Order 2 (Columns/Up Diagonal):}
\[
R(d_{11}) \leq 
R(d_{21}) \leq  
R(d_{12}) \leq 
R(d_{22}) \leq 
R(d_{13}) \leq 
R(d_{23}) \leq 
R(d_{14}) \leq 
R(d_{24}) \leq 
R(d_{15}) \leq 
R(d_{25}) 
\]
\textbf{Order 3 (Up-down Diagonal):}
\[
R(d_{11}) \leq 
R(d_{12}) \leq 
R(d_{21}) \leq  
R(d_{13}) \leq 
R(d_{22}) \leq 
R(d_{14}) \leq 
R(d_{23}) \leq 
R(d_{15}) \leq 
R(d_{24}) \leq 
R(d_{25}) 
\]
\textbf{Order 4 (Down-up Diagonal):}
\[
R(d_{11}) \leq 
R(d_{12}) \leq 
R(d_{21}) \leq  
R(d_{22}) \leq 
R(d_{13}) \leq 
R(d_{14}) \leq 
R(d_{23}) \leq 
R(d_{24}) \leq 
R(d_{15}) \leq 
R(d_{25}) 
\]
\textbf{Order 5 (Down Diagonal):}
\[
R(d_{11}) \leq 
R(d_{21}) \leq  
R(d_{12}) \leq 
R(d_{13}) \leq 
R(d_{22}) \leq 
R(d_{14}) \leq 
R(d_{23}) \leq 
R(d_{15}) \leq 
R(d_{24}) \leq 
R(d_{25}) 
\]
\begin{figure}[h!]
  \centering
  \includegraphics[width=0.8\linewidth]{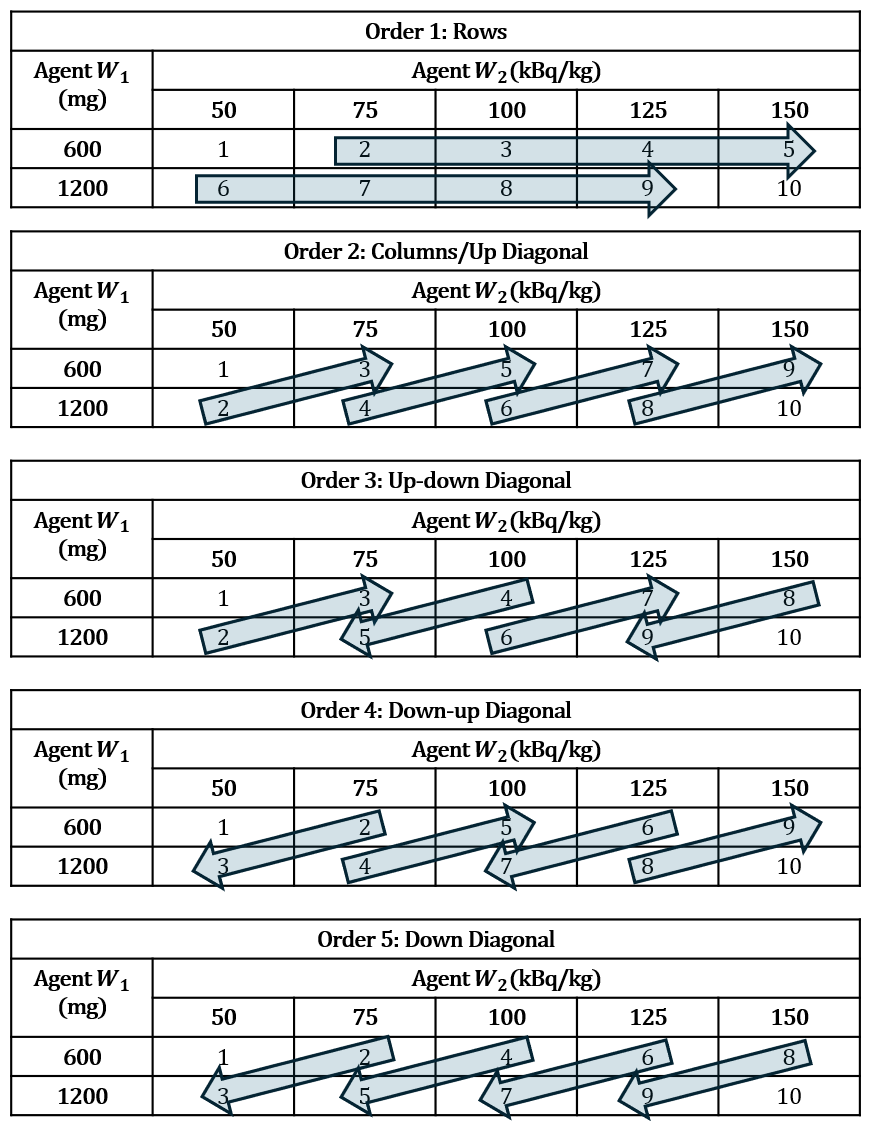}
\caption{Visualisation of the five partial orderings considered in the $2 \times 5$ grid.}
\label{fig:Vis2x5}
\end{figure}

\pagebreak

For the $3 \times 3$ grid:\\

\textbf{Order 1 (Rows):}
\[
R(d_{11}) \leq 
R(d_{12}) \leq 
R(d_{13}) \leq 
R(d_{21}) \leq  
R(d_{22}) \leq 
R(d_{23}) \leq 
R(d_{31}) \leq 
R(d_{32}) \leq 
R(d_{33}) 
\]

\textbf{Order 2 (Columns):}
\[
R(d_{11}) \leq 
R(d_{21}) \leq  
R(d_{31}) \leq 
R(d_{12}) \leq 
R(d_{22}) \leq 
R(d_{32}) \leq 
R(d_{13}) \leq 
R(d_{23}) \leq 
R(d_{33}) 
\]

\textbf{Order 3 (Down-up Diagonal):}
\[
R(d_{11}) \leq 
R(d_{21}) \leq  
R(d_{12}) \leq 
R(d_{13}) \leq 
R(d_{22}) \leq 
R(d_{31}) \leq 
R(d_{32}) \leq 
R(d_{23}) \leq 
R(d_{33}) 
\]

\textbf{Order 4 (Up-down Diagonal):}
\[
R(d_{11}) \leq 
R(d_{12}) \leq 
R(d_{21}) \leq  
R(d_{31}) \leq 
R(d_{22}) \leq 
R(d_{13}) \leq 
R(d_{23}) \leq 
R(d_{32}) \leq 
R(d_{33}) 
\]

\textbf{Order 5 (Up Diagonal):}
\[
R(d_{11}) \leq 
R(d_{12}) \leq 
R(d_{21}) \leq  
R(d_{13}) \leq 
R(d_{22}) \leq 
R(d_{31}) \leq 
R(d_{23}) \leq 
R(d_{32}) \leq 
R(d_{33}) 
\]

\textbf{Order 6 (Down Diagonal):}
\[
R(d_{11}) \leq 
R(d_{21}) \leq  
R(d_{12}) \leq 
R(d_{31}) \leq 
R(d_{22}) \leq 
R(d_{13}) \leq 
R(d_{32}) \leq 
R(d_{23}) \leq 
R(d_{33}). 
\]
\begin{figure}[h!]
  \centering
  \includegraphics[width=1\linewidth]{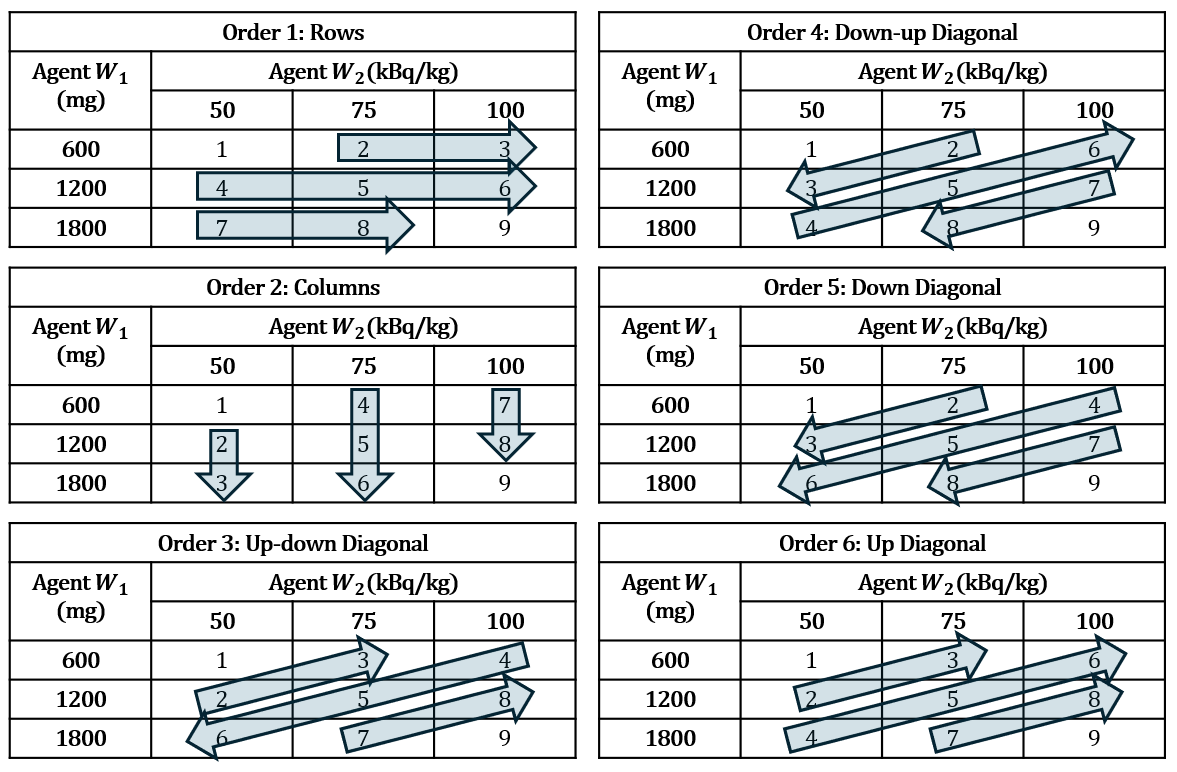}
\caption{Visualisation of the six partial orderings considered in the $3 \times 3$ grid.}
\label{fig:Vis3x3}
\end{figure}

The utility function uses two weights, $\omega_1$ and $\omega_2$, with $\omega_1$=0.33 and $\omega_2$=1.09, as used by Barnett et al. \cite{Barnett2024}. 
\pagebreak
\subsubsection{Joint TITE-BLRM} 
For each agent, the prior is elicited on the $\bm{\beta}$ in the same way as the Joint TITE-POCRM in Equation \ref{eq:mvn}, with values calibrated via the procedure outline in Section~\ref{sec:prior_cal}. Values of these hyper-parameters for each of the agents are given in Table~\ref{tab:hyp}.

The interaction parameter between activity and toxicity is given a vague prior,  $\psi \sim N(0,100)$. The interaction parameter between the agents $W_1$ and $W_2$ for both activity and toxicity is given a half-normal prior $\eta^{(T)}, \eta^{(A)} \sim HN(0,1)$.

The utility function uses the same weights as the Joint TITE-POCRM, $\omega_1$=0.33 and $\omega_2$=1.09.

\subsubsection{TITE-BOIN12}
The utility, $u$ has a $Beta(1,1)$ prior, with $\Psi_{Y^{(T)} Y^{(A)}}$ defined as: $\Psi_{01}=100$,  $\Psi_{00}=40$,  $\Psi_{10}=0$,  $\Psi_{11}=60$, as suggested by Zhou et al \cite{Zhou2022}.

\section{Likelihoods}

\subsection{Joint TITE-POCRM}
The $G_{P2}$ for activity and toxicity are related by a Gumbel Model considering all binary combinations of toxicity and activity outcomes:
\begin{align*}
\pi_{y^{(A)}_{\ell},{y^{(T)}_{\ell}}}(x^{(A)}_{\ell ,m^*_A},x^{(T)}_{\ell ,m^*_T}, \bm{\beta},\bm{w_{\ell}})= &(G_{P2}(x^{(A)}_{\ell ,m^*_A},w_{\ell}^{(A)},\bm{\beta^{(A)}}) )^{y^{(A)}_{\ell}}\\
& (1-G_{P2}(x^{(A)}_{\ell ,m^*_A},w_{\ell}^{(A)},\bm{\beta^{(A)}}) )^{1-{y^{(A)}_{\ell}} } \\ 
&(G_{P2}(x^{(T)}_{\ell ,m^*_T},w_{\ell}^{(T)},\bm{\beta^{(T)}} ))^{y^{(T)}_{\ell}}\\ 
 &(1-G_{P2}(x^{(T)}_{\ell ,m^*_T},w_{\ell}^{(T)},\bm{\beta^{(T)}}) )^{1-{y^{(T)}_{\ell}}} + \\ &(-1)^{{y^{(A)}_{\ell}} + {y^{(T)}_{\ell}}}(G_{P2}(x^{(A)}_{\ell ,m^*_A},w_{\ell}^{(A)},\bm{\beta^{(A)}}) )\\ 
 &(1-G_{P2}(x^{(A)}_{\ell ,m^*_A},w_{\ell}^{(A)},\bm{\beta^{(A)}}) ) \\ 
 &G_{P2}(x^{(T)}_{\ell ,m^*_T},w_{\ell}^{(T)},\bm{\beta^{(T)}}))\\  
 &(1-G_{P2}(x^{(T)}_{\ell ,m^*_T},w_{\ell}^{(T)},\bm{\beta^{(T)}}) ) \left( \frac{e^{\psi}-1}{e^{\psi}+1} \right),
\end{align*}
where $\pi_{y^{(A)}_{\ell},{y^{(T)}_{\ell}}}$ is the probability of observing each of the four combinations of binary activity outcomes, $\bm{\beta}=(\bm{\beta^{(A)}}, \bm{\beta^{(T)}})$ , $\bm{w_{\ell}}=(w_{\ell}^{(A)}, w_{\ell}^{(T)})$ and $\psi \in (-\infty , \infty)$ is a correlation parameter of the distribution, with positive values of $\psi$ corresponding to a positive correlation between activity and toxicity, negative values corresponding to a negative correlation, and larger absolute values of  $\psi$ corresponding to stronger correlations.\\ \\
The likelihood is then:
\begin{equation} \label{eq:LIK}
\mathcal{L}_{P2}(\bm{\beta}) = \prod_{\ell=1}^{n} \prod_{a=0}^{1} \prod_{b=0}^{1} \{\pi_{a,b}(x^{(A)}_{\ell ,m^*_A},x^{(T)}_{\ell ,m^*_T},\bm{\beta}, \bm{w_{\ell}})\}^{\mathbb{I}((Y^{(A)}_{\ell},Y^{(T)}_{\ell})=(a,b))},
\end{equation}
for patients $\ell=1, \ldots , n$.

\subsection{Joint TITE-BLRM}

The likelihood for the Joint TITE-BLRM is:
\begin{equation} \label{eq:LIK.BLRM}
\mathcal{L}_B(\bm{\beta}, \bm{\eta}, \psi) = \prod_{\ell=1}^{n} \prod_{a=0}^{1} \prod_{b=0}^{1} \{\pi_{a,b}(x_{\ell}^{W_1},x_{\ell}^{W_2},\bm{\beta}, \bm{\eta}, \psi, \bm{w_{\ell}})\}^{\mathbb{I}((Y^{(A)}_{\ell},Y^{(T)}_{\ell})=(a,b))},
\end{equation}

where 
\begin{align*}
\pi_{y^{(A)}_{\ell},{y^{(T)}_{\ell}}}&(x_{\ell}^{W_1},x_{\ell}^{W_2}, \bm{\beta}, \bm{\eta}, \psi, \bm{w_{\ell}})= \\
&(G_B(x_{\ell}^{W_1},x_{\ell}^{W_2},w_{\ell}^{(A)},\bm{\beta^{(W_1,A)}},\bm{\beta^{(W_2,A)}}, \eta^{(A)} )^{y^{(A)}_{\ell}}\\
&(1-G_B(x_{\ell}^{W_1},x_{\ell}^{W_2},w_{\ell}^{(A)},\bm{\beta^{(W_1,A)}},\bm{\beta^{(W_2,A)}}, \eta^{(A)} )^{1-{y^{(A)}_{\ell}} } \\ 
&(G_B(x_{\ell}^{W_1},x_{\ell}^{W_2},w_{\ell}^{(T)},\bm{\beta^{(W_1, T)}},\bm{\beta^{(W_2, T)}}, \eta^{(T)} )^{y^{(T)}_{\ell}} \\
& (1-G_B(x_{\ell}^{W_1},x_{\ell}^{W_2},w_{\ell}^{(T)},\bm{\beta^{(W_1, T)}},\bm{\beta^{(W_2, T)}} , \eta^{(T)})^{1-{y^{(T)}_{\ell}}} + \\ 
&(-1)^{{y^{(A)}_{\ell}} + {y^{(T)}_{\ell}}} (G_B(x_{\ell}^{W_1},x_{\ell}^{W_2},w_{\ell}^{(A)},\bm{\beta^{(W_1,A)}},\bm{\beta^{(W_2,A)}} , \eta^{(A)}) \\
&(1-G_B(x_{\ell}^{W_1},x_{\ell}^{W_2},w_{\ell}^{(A)},\bm{\beta^{(W_1,A)}},\bm{\beta^{(W_2,A)}}, \eta^{(A)}) \\ &G_B(x_{\ell}^{W_1},x_{\ell}^{W_2},w_{\ell}^{(T)},\bm{\beta^{(W_1, T)}},\bm{\beta^{(W_2, T)}}, \eta^{(T)})  \\
&(1-G_B(x_{\ell}^{W_1},x_{\ell}^{W_2},w_{\ell}^{(T)},\bm{\beta^{(W_1, T)}},\bm{\beta^{(W_2, T)}} , \eta^{(T)})) \left( \frac{e^{\psi}-1}{e^{\psi}+1} \right),
\end{align*}
where $\bm{\beta}=( \bm{\beta^{(W_1,A)}}, \bm{\beta^{(W_2,A)}}, \bm{\beta^{(W_1,T)}}, \bm{\beta^{(W_2,T)}})$ and $\bm{\eta}=(\eta^{(A)},\eta^{(T)})$

\section{Utility Scenarios}
The utility for each dose combination in each of the 36 scenarios for each method is given in the following tables.

Acceptable dose combinations are highlighted in \textit{italics}, good dose combinations are \uu{underlined} and correct dose combinations are highlighted in \textbf{boldface}.

\begin{landscape}

\begin{table}[h!]
{\scriptsize
\begin{tabular}{lllllllllllllllll}

&&  \multicolumn{5}{c}{A1}                                                                                                                                                    & \multicolumn{5}{c}{A2}                                                                                                                                                    & \multicolumn{5}{c}{A3}                                                                                                               \\ 
 & \multicolumn{1}{c}{\multirow{2}{*}{\begin{tabular}[c]{@{}c@{}}Agent $W_1$ \\ (mg)\end{tabular}}} & \multicolumn{15}{c}{Agent $W_2$ (kBq/kg)}  \\ \cline{3-17}                                      
                                        & \multicolumn{1}{c}{}                                                                                          & \multicolumn{1}{|c}{50}      
                                                                    & 75                            & 100                  & 125                           & \multicolumn{1}{l|}{150}                           & 50                            & 75                   & 100                           & 125                           & \multicolumn{1}{l|}{150}                           & 50     & 75     & 100                           & 125                           & \multicolumn{1}{l|}{150}                           \\ \cline{2-17} 
\multicolumn{1}{c}{\multirow{2}{*}{T1}} & \multicolumn{1}{|l|}{600}                                                                                      & \textit{0.190}                & \textit{0.277}                & \textit{0.314}       & \textit{0.451}                & \multicolumn{1}{l|}{{\ul \textit{0.484}}}          & \textit{0.290}                & {\ul \textit{0.297}} & {\ul \textit{0.344}}          & {\ul \textit{0.351}}          & \multicolumn{1}{l|}{{\ul \textit{\textbf{0.394}}}} & 0.050  & 0.057  & 0.084                         & \textit{0.151}                & \multicolumn{1}{l|}{{\ul \textit{0.234}}}          \\
\multicolumn{1}{c}{}                    & \multicolumn{1}{|l|}{1200}                                                                       & \textit{0.234}                & \textit{0.370}                & \textit{0.407}       & {\ul \textit{0.518}}          & \multicolumn{1}{l|}{{\ul \textit{\textbf{0.551}}}} & {\ul \textit{0.324}}          & {\ul \textit{0.330}} & {\ul \textit{0.377}}          & {\ul \textit{0.358}}          & \multicolumn{1}{l|}{{\ul \textit{0.381}}}          & 0.084  & 0.120  & {\ul \textit{0.207}}          & {\ul \textit{0.268}}          & \multicolumn{1}{l|}{{\ul \textit{\textbf{0.301}}}} \\ \cline{2-17} 
\multirow{2}{*}{T2}                     & \multicolumn{1}{|l|}{600}                                                                        & \textit{0.167}                & \textit{0.251}                & \textit{0.284}       & {\ul \textit{\textbf{0.401}}} & \multicolumn{1}{l|}{-0.672}                        & {\ul \textit{0.267}}          & {\ul \textit{0.271}} & {\ul \textit{\textbf{0.314}}} & {\ul \textit{0.301}}          & \multicolumn{1}{l|}{-0.762}                        & 0.027  & 0.031  & 0.054                         & {\ul \textit{\textbf{0.101}}} & \multicolumn{1}{l|}{-0.922}                        \\
                                        & \multicolumn{1}{|l|}{1200}                                                                       & -0.989                        & -0.855                        & -0.822               & -0.688                        & \multicolumn{1}{l|}{-0.638}                        & -0.899                        & -0.895               & -0.852                        & -0.848                        & \multicolumn{1}{l|}{-0.808}                        & -1.139 & -1.105 & -1.022                        & -0.938                        & \multicolumn{1}{l|}{-0.888}                        \\ \cline{2-17} 
\multirow{2}{*}{T3}                     & \multicolumn{1}{|l|}{600}                                                                        & \textit{0.184}                & \textit{0.274}                & \textit{0.301}       & {\ul \textit{\textbf{0.434}}} & \multicolumn{1}{l|}{-0.689}                        & {\ul \textit{0.284}}          & {\ul \textit{0.294}} & {\ul \textit{0.331}}          & {\ul \textit{\textbf{0.334}}} & \multicolumn{1}{l|}{-0.779}                        & 0.044  & 0.054  & 0.071                         & {\ul \textit{0.134}}          & \multicolumn{1}{l|}{-0.939}                        \\
                                        & \multicolumn{1}{|l|}{1200}                                                                       & \textit{0.217}                & {\ul \textit{0.360}}          & {\ul \textit{0.351}} & -0.622                        & \multicolumn{1}{l|}{-0.605}                        & {\ul \textit{0.307}}          & {\ul \textit{0.320}} & {\ul \textit{0.321}}          & -0.782                        & \multicolumn{1}{l|}{-0.775}                        & 0.067  & 0.110  & {\ul \textit{\textbf{0.151}}} & -0.872                        & \multicolumn{1}{l|}{-0.855}                        \\ \cline{2-17} 
\multirow{2}{*}{T4}                     & \multicolumn{1}{|l|}{600}                                                                        & {\ul \textit{0.167}}          & {\ul \textit{\textbf{0.234}}} & -0.872               & -0.755                        & \multicolumn{1}{l|}{-0.738}                        & {\ul \textit{\textbf{0.267}}} & {\ul \textit{0.254}} & -0.842                        & -0.855                        & \multicolumn{1}{l|}{-0.828}                        & 0.027  & 0.014  & -1.102                        & -1.055                        & \multicolumn{1}{l|}{-0.988}                        \\
                                        & \multicolumn{1}{|l|}{1200}                                                                       & {\ul \textit{0.151}}          & -0.839                        & -0.822               & -0.688                        & \multicolumn{1}{l|}{-0.638}                        & {\ul \textit{0.241}}          & -0.879               & -0.852                        & -0.848                        & \multicolumn{1}{l|}{-0.808}                        & 0.001  & -1.089 & -1.022                        & -0.938                        & \multicolumn{1}{l|}{-0.888}                        \\ \cline{2-17} 
\multirow{2}{*}{T5}                     & \multicolumn{1}{|l|}{600}                                                                        & {\ul \textit{\textbf{0.101}}} & -0.939                        & -0.905               & -0.772                        & \multicolumn{1}{l|}{-0.738}                        & {\ul \textit{\textbf{0.201}}} & -0.919               & -0.875                        & -0.872                        & \multicolumn{1}{l|}{-0.828}                        & -0.039 & -1.159 & -1.135                        & -1.072                        & \multicolumn{1}{l|}{-0.988                                            } \\
                                        & \multicolumn{1}{|l|}{1200}                                                                       & -0.972                        & -0.855                        & -0.822               & -0.688                        & \multicolumn{1}{l|}{-0.638}                        & -0.882                        & -0.895               & -0.852                        & -0.848                        & \multicolumn{1}{l|}{-0.808}                        & -1.122 & -1.105 & -1.022                        & -0.938                        & \multicolumn{1}{l|}{-0.888}                                   \\ \cline{2-17} 
\multirow{2}{*}{T6}                     & \multicolumn{1}{|l|}{600}                                                                        & -1.022                        & -0.922                        & -0.905               & -0.755                        & \multicolumn{1}{l|}{-0.738}                        & -0.922                        & -0.902               & -0.875                        & -0.855                        & \multicolumn{1}{l|}{-0.828}                        & -1.162 & -1.142 & -1.135                        & -1.055                        & \multicolumn{1}{l|}{-0.988}                        \\
                                        & \multicolumn{1}{|l|}{1200}                                                                       & -0.972                        & -0.822                        & -0.805               & -0.655                        & \multicolumn{1}{l|}{-0.638}                        & -0.882                        & -0.862               & -0.835                        & -0.815                        & \multicolumn{1}{l|}{-0.808}                        & -1.122 & -1.072 & -1.005                        & -0.905                        & \multicolumn{1}{l|}{-0.888}                        \\ \cline{2-17} 

\end{tabular}}
\caption{POCRM/BLRM utility}
\end{table}

\begin{table}[h!]
{\scriptsize
\begin{tabular}{lllllllllllllllll}

&&  \multicolumn{5}{c}{A4}                                                                                                                                                    & \multicolumn{5}{c}{A5}                                                                                                                                                    & \multicolumn{5}{c}{A6}                                                                                                               \\ 
 & \multicolumn{1}{c}{\multirow{2}{*}{\begin{tabular}[c]{@{}c@{}}Agent $W_1$ \\ (mg)\end{tabular}}} & \multicolumn{15}{c}{Agent $W_2$ (kBq/kg)}  \\ \cline{3-17}                                      
                                        & \multicolumn{1}{c}{}                                                                                          & \multicolumn{1}{|c}{50}      
                                                                    & 75                            & 100                  & 125                           & \multicolumn{1}{l|}{150}                           & 50                            & 75                   & 100                           & 125                           & \multicolumn{1}{l|}{150}                           & 50     & 75     & 100                           & 125                           & \multicolumn{1}{l|}{150}                           \\ \cline{2-17} 
\multicolumn{1}{c}{\multirow{2}{*}{T1}} & \multicolumn{1}{|l|}{600}                                                                        & 0.040  & \textit{0.177}                & \textit{0.264}       & \textit{0.351}                & \multicolumn{1}{l|}{{\ul \textit{0.434}}}          & 0.090                         & 0.097                & 0.104                         & 0.111                & \multicolumn{1}{l|}{0.114}                         & 0.090  & 0.077  & 0.064  & 0.051  & \multicolumn{1}{l|}{0.034}                         \\
\multicolumn{1}{c}{}                    & \multicolumn{1}{|l|}{1200}                                                                       & 0.084  & \textit{0.220}                & \textit{0.307}       & {\ul \textit{0.368}}          & \multicolumn{1}{l|}{{\ul \textit{\textbf{0.451}}}} & \textit{0.184}                & \textit{0.270}       & \textit{0.357}                & {\ul \textit{0.418}} & \multicolumn{1}{l|}{{\ul \textit{\textbf{0.501}}}} & 0.084  & 0.070  & 0.057  & 0.018  & \multicolumn{1}{l|}{{\ul \textit{\textbf{0.101}}}} \\ \cline{2-17} 
\multirow{2}{*}{T2}                     & \multicolumn{1}{|l|}{600}                                                                        & 0.017  & \textit{0.151}                & {\ul \textit{0.234}} & {\ul \textit{\textbf{0.301}               }} & \multicolumn{1}{l|}{-0.722}                        & 0.067                         & 0.071                & 0.074                         & 0.061                & \multicolumn{1}{l|}{-1.042}                        & 0.067  & 0.051  & 0.034  & 0.001  & \multicolumn{1}{l|}{-1.122}                        \\
                                        & \multicolumn{1}{|l|}{1200}                                                                       & -1.139 & -1.005                        & -0.922               & -0.838                        & \multicolumn{1}{l|}{-0.738}                        & -1.039                        & -0.955               & -0.872                        & -0.788               & \multicolumn{1}{l|}{-0.688}                        & -1.139 & -1.155 & -1.172 & -1.188 & \multicolumn{1}{l|}{-1.088}                        \\ \cline{2-17} 
\multirow{2}{*}{T3}                     & \multicolumn{1}{|l|}{600}                                                                        & 0.034  & \textit{0.174}                & {\ul \textit{0.251}} & {\ul \textit{\textbf{0.334}}} & \multicolumn{1}{l|}{-0.739}                        & 0.084                         & 0.094                & 0.091                         & 0.094                & \multicolumn{1}{l|}{-1.059}                        & 0.084  & 0.074  & 0.051  & 0.034  & \multicolumn{1}{l|}{-1.139}                        \\
                                        & \multicolumn{1}{|l|}{1200}                                                                       & 0.067  & \textit{0.210}                & {\ul \textit{0.251}} & -0.772                        & \multicolumn{1}{l|}{-0.705}                        & \textit{0.167}                & {\ul \textit{0.260}} & {\ul \textit{\textbf{0.301}}} & -0.722               & \multicolumn{1}{l|}{-0.655}                        & 0.067  & 0.060  & 0.001  & -1.122 & \multicolumn{1}{l|}{-1.055}                        \\ \cline{2-17} 
\multirow{2}{*}{T4}                     & \multicolumn{1}{|l|}{600}                                                                        & 0.017  & {\ul \textit{\textbf{0.134}}} & -0.922               & -0.855                        & \multicolumn{1}{l|}{-0.788}                        & 0.067                         & 0.054                & -1.082                        & -1.095               & \multicolumn{1}{l|}{-1.108}                        & 0.067  & 0.034  & -1.122 & -1.155 & \multicolumn{1}{l|}{-1.188}                        \\
                                        & \multicolumn{1}{|l|}{1200}                                                                       & 0.001  & -0.989                        & -0.922               & -0.838                        & \multicolumn{1}{l|}{-0.738}                        & {\ul \textit{\textbf{0.101}}} & -0.939               & -0.872                        & -0.788               & \multicolumn{1}{l|}{-0.688}                        & 0.001  & -1.139 & -1.172 & -1.188 & \multicolumn{1}{l|}{-1.088}                        \\ \cline{2-17} 
\multirow{2}{*}{T5}                     & \multicolumn{1}{|l|}{600}                                                                        & -0.049 & -1.039                        & -0.955               & -0.872                        & \multicolumn{1}{l|}{-0.788}                        & 0.001                         & -1.119               & -1.115                        & -1.112               & \multicolumn{1}{l|}{-1.108}                        & 0.001  & -1.139 & -1.155 & -1.172 & \multicolumn{1}{l|}{-1.188}                        \\
                                        & \multicolumn{1}{|l|}{1200}                                                                       & -1.122 & -1.005                        & -0.922               & -0.838                        & \multicolumn{1}{l|}{-0.738}                        & -1.022                        & -0.955               & -0.872                        & -0.788               & \multicolumn{1}{l|}{-0.688}                        & -1.122 & -1.155 & -1.172 & -1.188 & \multicolumn{1}{l|}{-1.088}                        \\ \cline{2-17} 
\multirow{2}{*}{T6}                     & \multicolumn{1}{|l|}{600}                                                                        & -1.172 & -1.022                        & -0.955               & -0.855                        & \multicolumn{1}{l|}{-0.788}                        & -1.122                        & -1.102               & -1.115                        & -1.095               & \multicolumn{1}{l|}{-1.108}                        & -1.122 & -1.122 & -1.155 & -1.155 & \multicolumn{1}{l|}{-1.188}                        \\
                                        & \multicolumn{1}{|l|}{1200}                                                                       & -1.122 & -0.972                        & -0.905               & -0.805                        & \multicolumn{1}{l|}{-0.738}                        & -1.022                        & -0.922               & -0.855                        & -0.755               & \multicolumn{1}{l|}{-0.688}                        & -1.122 & -1.122 & -1.155 & -1.155 & \multicolumn{1}{l|}{-1.088}                        \\ \cline{2-17} 
\end{tabular}}
\caption{POCRM/BLRM utility}
\end{table}

\begin{table}[h!]
{\scriptsize
\begin{tabular}{lllllllllllllllll}
   
&&  \multicolumn{5}{c}{A1}                                                                                                                                                    & \multicolumn{5}{c}{A2}                                                                                                                                                    & \multicolumn{5}{c}{A3}                                                                                                               \\ 
 & \multicolumn{1}{c}{\multirow{2}{*}{\begin{tabular}[c]{@{}c@{}}Agent $W_1$ \\ (mg)\end{tabular}}} & \multicolumn{15}{c}{Agent $W_2$ (kBq/kg)}  \\ \cline{3-17}                                      
                                        & \multicolumn{1}{c}{}                                                                                          & \multicolumn{1}{|c}{50}      
                                                                    & 75                            & 100                  & 125                           & \multicolumn{1}{l|}{150}                           & 50                            & 75                   & 100                           & 125                           & \multicolumn{1}{l|}{150}                           & 50     & 75     & 100                           & 125                           & \multicolumn{1}{l|}{150}                           \\ \cline{2-17} 
\multicolumn{1}{c}{\multirow{2}{*}{T1}} & \multicolumn{1}{|l|}{600}                                                                        & \textit{50.8}              & \textit{55.2}              & \textit{56.6} & {\ul \textit{64}}          & \multicolumn{1}{l|}{{\ul \textit{65}}}          & {\ul \textit{56.8}}        & {\ul \textit{56.4}} & {\ul \textit{58.4}}          & {\ul \textit{58}}   & \multicolumn{1}{l|}{{\ul \textit{59.6}}} & 42.4 & 42   & 42.8                & \textit{46}                & \multicolumn{1}{l|}{{\ul \textit{50}}} \\
\multicolumn{1}{c}{}                    & \multicolumn{1}{|l|}{1200}                                                                       & \textit{53}                & \textit{60.4}              & \textit{61.8} & {\ul \textit{66}}          & \multicolumn{1}{l|}{{\ul \textit{\textbf{67}}}} & {\ul \textit{58.4}}        & {\ul \textit{58}}   & {\ul \textit{\textbf{60}}}   & {\ul \textit{56.4}} & \multicolumn{1}{l|}{{\ul \textit{56.8}}} & 44   & 45.4 & {\ul \textit{49.8}} & {\ul \textit{51}}          & \multicolumn{1}{l|}{{\ul \textbf{52}}} \\ \cline{2-17} 
\multirow{2}{*}{T2}                     & \multicolumn{1}{|l|}{600}                                                                        & \textit{48}                & \textit{52}                & {\ul 53}      & {\ul \textit{\textbf{58}}} & \multicolumn{1}{l|}{57}                         & {\ul \textit{54}}          & {\ul \textit{53.2}} & {\ul \textit{\textbf{54.8}}} & {\ul \textit{52}}   & \multicolumn{1}{l|}{51.6}                & 39.6 & 38.8 & 39.2                & {\ul \textit{\textbf{40}}} & \multicolumn{1}{l|}{42}                \\
                                        & \multicolumn{1}{|l|}{1200}                                                                       & 37                         & 44                         & 45            & 52                         & \multicolumn{1}{l|}{55}                         & 42.4                       & 41.6                & 43.2                         & 42.4                & \multicolumn{1}{l|}{44.8}                & 28   & 29   & 33                  & 37                         & \multicolumn{1}{l|}{40}                \\ \cline{2-17} 
\multirow{2}{*}{T3}                     & \multicolumn{1}{|l|}{600}                                                                        & \textit{50}                & \textit{54.8}              & 55            & {\ul \textit{\textbf{62}}} & \multicolumn{1}{l|}{55}                         & {\ul \textit{56}}          & {\ul \textit{56}}   & {\ul \textit{\textbf{56.8}}} & {\ul \textit{56}}   & \multicolumn{1}{l|}{49.6}                & 41.6 & 41.6 & 41.2                & {\ul \textit{\textbf{44}}} & \multicolumn{1}{l|}{40}                \\
                                        & \multicolumn{1}{|l|}{1200}                                                                       & \textit{51}                & {\ul \textit{59.2}}        & 55            & 60                         & \multicolumn{1}{l|}{59}                         & {\ul \textit{56.4}}        & {\ul \textit{56.8}} & {\ul \textit{53.2}}          & 50.4                & \multicolumn{1}{l|}{48.8}                & 42   & 44.2 & {\ul \textit{43}}   & 45                         & \multicolumn{1}{l|}{44}                \\ \cline{2-17} 
\multirow{2}{*}{T4}                     & \multicolumn{1}{|l|}{600}                                                                        & \textit{48}                & {\ul \textit{\textbf{50}}} & 45            & 50                         & \multicolumn{1}{l|}{49}                         & {\ul \textit{\textbf{54}}} & {\ul \textit{51.2}} & 46.8                         & 44                  & \multicolumn{1}{l|}{43.6}                & 39.6 & 36.8 & 31.2                & 32                         & \multicolumn{1}{l|}{34}                \\
                                        & \multicolumn{1}{|l|}{1200}                                                                       & 43                         & 46                         & 45            & 52                         & \multicolumn{1}{l|}{55}                         & \textit{48.4}              & 43.6                & 43.2                         & 42.4                & \multicolumn{1}{l|}{44.8}                & 34   & 31   & 33                  & 37                         & \multicolumn{1}{l|}{40}                \\ \cline{2-17} 
\multirow{2}{*}{T5}                     & \multicolumn{1}{|l|}{600}                                                                        & {\ul \textit{\textbf{40}}} & 40                         & 41            & 48                         & \multicolumn{1}{l|}{49}                         & {\ul \textit{\textbf{46}}} & 41.2                & 42.8                         & 42                  & \multicolumn{1}{l|}{43.6}                & 31.6 & 26.8 & 27.2                & 30                         & \multicolumn{1}{l|}{34}                \\
                                        & \multicolumn{1}{|l|}{1200}                                                                       & 39                         & 44                         & 45            & 52                         & \multicolumn{1}{l|}{55}                         & 44.4                       & 41.6                & 43.2                         & 42.4                & \multicolumn{1}{l|}{44.8}                & 30   & 29   & 33                  & 37                         & \multicolumn{1}{l|}{40}                \\ \cline{2-17} 
\multirow{2}{*}{T6}                     & \multicolumn{1}{|l|}{600}                                                                        & 36                         & 42                         & 41            & 50                         & \multicolumn{1}{l|}{49}                         & 42                         & 43.2                & 42.8                         & 44                  & \multicolumn{1}{l|}{43.6}                & 27.6 & 28.8 & 27.2                & 32                         & \multicolumn{1}{l|}{34}                \\
                                        & \multicolumn{1}{|l|}{1200}                                                                       & 39                         & 48                         & 47            & 56                         & \multicolumn{1}{l|}{55}                         & 44.4                       & 45.6                & 45.2                         & 46.4                & \multicolumn{1}{l|}{44.8}                & 30   & 33   & 35                  & 41                         & \multicolumn{1}{l|}{40}                \\ \cline{2-17} 
\end{tabular}}
\caption{BOIN utility}
\end{table}

\begin{table}[h!]
{\scriptsize
\begin{tabular}{lllllllllllllllll}

&&  \multicolumn{5}{c}{A4}                                                                                                                                                    & \multicolumn{5}{c}{A5}                                                                                                                                                    & \multicolumn{5}{c}{A6}                                                                                                               \\ 
 & \multicolumn{1}{c}{\multirow{2}{*}{\begin{tabular}[c]{@{}c@{}}Agent $W_1$ \\ (mg)\end{tabular}}} & \multicolumn{15}{c}{Agent $W_2$ (kBq/kg)}  \\ \cline{3-17}                                      
                                        & \multicolumn{1}{c}{}                                                                                          & \multicolumn{1}{|c}{50}      
                                                                    & 75                            & 100                  & 125                           & \multicolumn{1}{l|}{150}                           & 50                            & 75                   & 100                           & 125                           & \multicolumn{1}{l|}{150}                           & 50     & 75     & 100                           & 125                           & \multicolumn{1}{l|}{150}                           \\ \cline{2-17} 
\multicolumn{1}{c}{\multirow{2}{*}{T1}} & \multicolumn{1}{|l|}{600}                                                                        & 41.8 & \textit{49.2}              & \textit{53.6}     & {\ul \textit{58}}          & \multicolumn{1}{l|}{{\ul \textit{\textbf{62}}}} & 44.8                       & 44.4                & 44                         & 43.6              & \multicolumn{1}{l|}{42.8}                       & 44.8 & 43.2 & 41.6 & 40  & \multicolumn{1}{l|}{38}                         \\
\multicolumn{1}{c}{}                    & \multicolumn{1}{|l|}{1200}                                                                       & 44   & \textit{51.4}              & \textit{55.8}     & {\ul \textit{57}}          & \multicolumn{1}{l|}{{\ul \textit{61}}}          & \textit{50}                & \textit{54.4}       & \textit{58.8}              & {\ul \textit{60}} & \multicolumn{1}{l|}{{\ul \textit{\textbf{64}}}} & 44   & 42.4 & 40.8 & 36  & \multicolumn{1}{l|}{{\ul \textit{\textbf{40}}}} \\ \cline{2-17} 
\multirow{2}{*}{T2}                     & \multicolumn{1}{|l|}{600}                                                                        & 39   & \textit{46}                & {\ul \textit{50}} & {\ul \textbf{52}}          & \multicolumn{1}{l|}{54}                         & 42                         & 41.2                & 40.4                       & 37.6              & \multicolumn{1}{l|}{34.8}                       & 42   & 40   & 38   & 34  & \multicolumn{1}{l|}{30}                         \\
                                        & \multicolumn{1}{|l|}{1200}                                                                       & 28   & 35                         & 39                & 43                         & \multicolumn{1}{l|}{49}                         & 34                         & 38                  & 42                         & 46                & \multicolumn{1}{l|}{52}                         & 28   & 26   & 24   & 22  & \multicolumn{1}{l|}{28}                         \\ \cline{2-17} 
\multirow{2}{*}{T3}                     & \multicolumn{1}{|l|}{600}                                                                        & 41   & \textit{48.8}              & \textit{52}       & {\ul \textit{\textbf{56}}} & \multicolumn{1}{l|}{52}                         & 44                         & 44                  & 42.4                       & 41.6              & \multicolumn{1}{l|}{32.8}                       & 44   & 42.8 & 40   & 38  & \multicolumn{1}{l|}{28}                         \\
                                        & \multicolumn{1}{|l|}{1200}                                                                       & 42   & \textit{50.2}              & {\ul \textit{49}} & 51                         & \multicolumn{1}{l|}{53}                         & \textit{48}                & {\ul \textit{53.2}} & {\ul \textit{\textbf{52}}} & 54                & \multicolumn{1}{l|}{56}                         & 42   & 41.2 & 34   & 30  & \multicolumn{1}{l|}{32}                         \\ \cline{2-17} 
\multirow{2}{*}{T4}                     & \multicolumn{1}{|l|}{600}                                                                        & 39   & {\ul \textit{\textbf{44}}} & 42                & 44                         & \multicolumn{1}{l|}{46}                         & 42                         & 39.2                & 32.4                       & 29.6              & \multicolumn{1}{l|}{26.8}                       & 42   & 38   & 30   & 26  & \multicolumn{1}{l|}{22}                         \\
                                        & \multicolumn{1}{|l|}{1200}                                                                       & 34   & 37                         & 39                & 43                         & \multicolumn{1}{l|}{49}                         & {\ul \textit{\textbf{40}}} & 40                  & 42                         & 46                & \multicolumn{1}{l|}{52}                         & 34   & 28   & 24   & 22  & \multicolumn{1}{l|}{28}                         \\ \cline{2-17} 
\multirow{2}{*}{T5}                     & \multicolumn{1}{|l|}{600}                                                                        & 31   & 34                         & 38                & 42                         & \multicolumn{1}{l|}{46}                         & 34                         & 29.2                & 28.4                       & 27.6              & \multicolumn{1}{l|}{26.8}                       & 34   & 28   & 26   & 24  & \multicolumn{1}{l|}{22}                         \\
                                        & \multicolumn{1}{|l|}{1200}                                                                       & 30   & 35                         & 39                & 43                         & \multicolumn{1}{l|}{49}                         & 36                         & 38                  & 42                         & 46                & \multicolumn{1}{l|}{52}                         & 30   & 26   & 24   & 22  & \multicolumn{1}{l|}{28}                         \\ \cline{2-17} 
\multirow{2}{*}{T6}                     & \multicolumn{1}{|l|}{600}                                                                        & 27   & 36                         & 38                & 44                         & \multicolumn{1}{l|}{46}                         & 30                         & 31.2                & 28.4                       & 29.6              & \multicolumn{1}{l|}{26.8}                       & 30   & 30   & 26   & 26  & \multicolumn{1}{l|}{22}                         \\
                                        & \multicolumn{1}{|l|}{1200}                                                                       & 30   & 39                         & 41                & 47                         & \multicolumn{1}{l|}{49}                         & 36                         & 42                  & 44                         & 50                & \multicolumn{1}{l|}{52}                         & 30   & 30   & 26   & 26  & \multicolumn{1}{l|}{28}                         \\ \cline{2-17} 
\end{tabular}}
\caption{BOIN utility}
\end{table}

\end{landscape}

\section{Non-parametric Benchmark}
In order to give the results for each of the scenarios a relevant interpretation, they are compared to the non-parametric benchmark for unknown orderings \cite{Mozgunov2021a}.

For this work, we have also made the modification to this benchmark to restrict selections based on admissibility, so that the comparison of results are meaningful. For exact details of the benchmark for unknown ordering, we refer the reader to the original manuscript by Mozgunov et al \cite{Mozgunov2021a}. In summary, like the traditional non-parametric benchmark, the complete information for all patients (the maximum sample size) on all doses is considered in order to find the dose that is closest to the target for some criterion. In the case of unknown ordering, this benchmark assesses the likelihood of each ordering, given the complete information on each patient. Since in the implementation of the Joint TITE-POCRM and Joint TITE-BLRM, the admissibility of each dose combination is an important feature in the recommendation of dose combination, this should also be included in the benchmark. Therefore, the same admissibility criteria that is applied to the dose combinations for each of the applied methods based on their models is applied to the complete information. In this way, the benchmark is also able to recommend "No admissible dose"  when none of the doses are admissible based on the complete information.